\documentclass[superscriptaddress,showkeys,prd,nofootinbib,floatfix,preprintnumbers]{revtex4-1}

\usepackage{multirow}
\usepackage{amsfonts,amsmath,amssymb} 
\usepackage{graphicx,graphics,color}
\usepackage{gensymb}
\usepackage{longtable}
\usepackage{bbding}
\usepackage{subfigure}
\usepackage{xcolor}
\usepackage{hyperref}
\usepackage[normalem]{ulem}
\hypersetup{
    colorlinks=true,
    linkcolor=blue}

       \def\jcap{J.
Cosm. Astropart. Phys.}

\begin{document}




\title{A new Parametrization for Bulk Viscosity Cosmology \\
as Extension of the $\Lambda$CDM Model}

\author{Gabriel G\'omez}
\email{gabriel.gomez.d@usach.cl}
\affiliation{Departamento de F\'isica, Universidad de Santiago de Chile,\\Avenida V\'ictor Jara 3493, Estaci\'on Central, 9170124, Santiago, Chile}

\author{Guillermo Palma}
\email{guillermo.palma@usach.cl}
\affiliation{Departamento de F\'isica, Universidad de Santiago de Chile,\\Avenida V\'ictor Jara 3493, Estaci\'on Central, 9170124, Santiago, Chile}

\author{Esteban Gonz\'alez}
\email{esteban.gonzalez@ucn.cl}
\affiliation{Departamento de Física, Universidad Católica del Norte, Avenida Angamos 0610, Casilla 1280, Antofagasta, Chile}

\author{\'Angel Rinc\'on}
\email{angel.rincon.r@usach.cl}
\affiliation{Departamento de F\'isica, Universidad de Santiago de Chile,\\Avenida V\'ictor Jara 3493, Estaci\'on Central, 9170124, Santiago, Chile}

\author{Norman Cruz}
\email{norman.cruz@usach.cl}
\affiliation{Departamento de F\'isica, Universidad de Santiago de Chile,\\Avenida V\'ictor Jara 3493, Estaci\'on Central, 9170124, Santiago, Chile}


\date{\today}

\begin{abstract}
Bulk viscosity in cold dark matter is an appealing feature that introduces distinctive phenomenological effects in the cosmological setting as compared to the $\Lambda$CDM model. Under this view, we propose a general parametrization of the bulk viscosity of the form $\xi\sim  H^{1-2s} \rho_{m}^{s}$, Some advantages of this novel parametrization are: first, it allows to write the resulting equations of cosmological evolution in the form of an autonomous system for any value of $s$, so a general treatment of the fixed points and stability can be done, and second, the bulk viscosity effect is consistently handled so that it naturally turns off when matter density vanishes. As a main result we find, based on detailed dynamical system analysis,  one-parameter family of de-Sitter-like asymptotic solutions with non-vanishing bulk viscosity coefficient during different cosmological periods. 
Numerical computations are performed jointly along with analytical phase space analysis in order to assess more quantitatively the bulk viscosity effect on the cosmological background evolution. Finally, as a first contact with observation we derive constraints on the free parameters of some bulk viscosity models with specific $s$-exponents from Supernovae Ia and observations of the Hubble parameter, by performing a Bayesian statistical analysis thought the Markov Chain Monte Carlo method.
 
\end{abstract}

\pacs{Valid PACS appear here}
\maketitle

\section{Introduction}

The $\Lambda$CDM model describes a Universe with a dark energy (DE) component modeled by a positive cosmological constant, which drives the recent accelerated expansion~\cite{Perlmutter:1998np}, and pressureless fluid representing the up to date unknown cold dark matter (DM) component, responsible for the structure formation in the Universe. This model has been very successful to fit very well the cosmological data \cite{Planck2018,WMAP2013, BAO2017,OHD2021}. Nevertheless, currently many tension are challenging the physics behind this model, such as measurements of the Hubble parameter at the current time, $H_{0}$, which exhibit a discrepancy of $4.4\sigma$ between the measurements obtained from Planck CMB and the locally measurements obtained by A. G. Riess \textit{et al.} \cite{Riess:2019cxk}.
Other tensions are the measurements of $\sigma_{8}-\Omega_{m}$ (where $\sigma_{8}$ is the r.m.s. fluctuations of perturbations at $8h^{-1}$~Mpc scale) coming from large scale structure (LSS) observations and the extrapolated from Planck CMB (dependent on the $\Lambda$CDM model) \cite{remedyforplankanlssdata,sigma8}, and the results from the experiment EDGES to detect the global absorption signal of 21 cm line during the dark ages, which reveal an excess of radiation in the reionization epoch that is not predicted by the $\Lambda$CDM model, specifically at $z \approx 17$ \cite{linia21cm}.

%
%
One approach used to attempt to overcome some of the mentioned problems is the inclusion of viscosity in the cosmological fluids in order to have a more realistic description of their nature beyond the perfect fluid idealization. For example, in  \cite{tensionH0,Wilson:2006gf,BulktensionH}, the authors address the $H_{0}$ tension as an important guidance to construct new cosmological models with viscous/inhomogeneous fluids. Also, in \cite{remedyforplankanlssdata} it is shown that the $\sigma_{8}-\Omega_{m}$ tension can be alleviated if one assumes a small amount of viscosity in the DM component; even more, the excess of radiation observed by EDGES experiment is explain in \cite{linea21cmH} by considering a viscous nature in DM. Nevertheless, due to the negative pressure that characterizes dissipative process in cosmic fluids, several authors have investigated the late time acceleration of the Universe as a pure effect of the bulk viscosity  \cite{paperprofeAccelerated,Bulk1,Bulk2,Bulk3,Bulk4,Bulk5,Bulk6,Bulk7,Bulk8,Exact,Testing,Almada1}, as an alternative mechanism to the one provided by the cosmological constant. 

For a homogeneous and isotropic universe, the dissipative process can be characterized only by bulk viscosity, which in the cosmic evolution has appealing effects \cite{Padmanabhan:1987dg,Zimdahl:2000zm,Wilson:2006gf,LandauandLifshitz} and from the macroscopic point of view can be interpreted as the existence of slow processes to restore the equilibrium state. Some authors have proposed that bulk viscosity may be the result of non-conserving particle interactions \cite{Cosmologycreation} or it could be the result of different cooling rates for the components of the cosmic medium  \cite{cooling1,cooling2,Zimdahl:1996fj}. In addition, many observational properties of disk galaxies can be described by a dissipative DM component \cite{foot2015dissipative,foot2016solving}. At perturbative level, a viscous fluid description is an accurate approach for extending the description of cosmological perturbations into a non-linear regime \cite{blas2015large}. In this same direction several works have investigated the perturbative effects of viscous DM models in the structure formation of the Universe \cite{remedyforplankanlssdata,Velten:2014xca,Barbosa:2017ojt,Kunz:2016yqy,Floerchinger:2016hja}. The inclusion of viscosity has been also investigated at early times aiming at describing the primordial inflationary period \cite{Gron:1990ew,Padmanabhan:1987dg} and, on the other hand, to evaluate the rate of cosmological entropy production and its role in the survival of protogalaxies \cite{weinberg1971entropy}.

On the other hand, there are several microscopic models to explain how bulk viscosity could arise in cosmological scenarios. Among them, we mention the inclusion of self-interacting scalar fields to describe dark energy (see \cite{Gagnon:2011id}), which gives rise to a contribution linearly proportional to the Hubble parameter to the fluid pressure, as in Eckart’s theory but in a more general setup within thermal field theory, where the viscosity coefficient becomes dynamical. Still, an astringent and throughout analysis of the assumptions for the validity of the hydrodynamical description used, including the effects of cavitation, is still missing. 
Also from a microscopic point of view, the relation between particle creation and bulk viscosity in the early universe is discussed in \cite{Brevik:1996ca,Murphy:1973zz,Hu:1982uea,Eshaghi:2015tqa}, which plays an important role in the inflationary viscous model \cite{Bamba:2015sxa}. In addition to the discussion made by \cite{Gagnon:2011id}, a different microscopic model that considers a bulk viscosity induced by DM annihilation is discussed in \cite{Wilson:2006gf, Mathews:2008hk}, suited for the late time accelerated expansion.
The kinetic theory formalism has been also implemented to describe the viscous effect within self-interacting DM models \cite{Atreya:2017pny, Natwariya:2019fif}.
In the context of neutralino CDM, an energy dissipation from the CDM fluid to the radiation fluid is manifested in a collisional damping mechanism during the kinetic decoupling \cite{Hofmann:2001bi}. The examples mentioned above highlight the importance of considering various dissipative processes and their potential effects on the cosmic fluid's dynamics. However, due to the lack of a general accepted model to include a microscopical motivated bulk viscosity, we propose instead a general effective parametrization that encompasses a wide class of possible models for the bulk viscosity, which allows us to describe a richer cosmological dynamics beyond the standard $\Lambda$CDM model.

Previous considerations indicate that viscous effect cannot be discarded at late times \cite{Oditsov3}, where the unidentified DM component is an essential protagonist, playing an important role from galactic dynamics to the formation of large scale structures in the Universe. To describe viscous cosmological models it is needed a theory of relativistic non-perfect fluids out of equilibrium. Under this framework, Eckart was the first to propose such a theory \cite{Eckart} with a similar approach proposed by Landau and Lifshitz \cite{LandauandLifshitz}. Nevertheless, it was shown later in \cite{Eckart,Muller} that the Eckart’s theory is a non-causal theory. Subsequently and following the same spirit, Israel and Stewart (IS) in \cite{I.S.1,I.S.2} introduced the corresponding fully causal version
%
which reduces to Eckart's theory when the relaxation time for the bulk viscous effects are negligible \cite{Dissipativecosmology}. 

Based on 
  i) the richness of the physics behind the bulk viscosity, 
 ii) the wide range of parameterizations proposed for the bulk viscosity, and finally, 
iii) the observational implications in the cosmic evolution,
we investigate in this paper a more involved parameterization including simultaneously the effects of both: first, the Hubble parameter, and second, the dark matter energy density. Our approach will be implemented within a viscous $\Lambda$CDM model.

%
%
Thus, the aim of this paper is to investigate the behavior of a concrete parameterization for the bulk viscosity of the form
\begin{equation}
\xi\sim H^{1-2s} \rho_{m}^{s},
\end{equation}
by means of detailed dynamical system analysis. Albeit non-trivial, the goodness of this novel functional form is that the bulk viscosity is consistently handled so that it naturally turns off when matter density vanishes.

Demanding solely complete cosmological dynamics, that is, from radiation era up to the present time, gives a relevant information about what particular parameterization for the bulk viscosity is suitable and sets, therefore, plausible dissipative cosmological models that must be subjected to the careful scrutiny requested for observational constraints. We perform this analysis in the framework of Eckart's theory as a first approximation to the study of relativistic non-perfect fluids. Thus, the dynamical analysis will indicate, as a first inquiry, what the values of the model parameters are of the proposed parametrization that leads to a successful background cosmic evolution, and this information is used \textit{a priori} to constrain the available parameter space with the aid of Supernovae Ia (SNe Ia) data and observational Hubble parameter data (OHD). 

The present paper is organized as follows: After a concise introduction, we summarize a few essential features of bulk viscosity and some explored parameterizations in dissipative cosmological models, discussing also the study of their behavior in the framework of dynamical analysis, in section II. Then, in section III, we present the model to be studied, as well as the concrete and new parameterization of the bulk viscosity that we will explore. Subsequently, in section IV, we perform a dynamical system analysis to set the stability conditions of the fixed points. We summarize our main results in tables I and II for concrete values of the exponent $s$.
A comprehensive analytical treatment is provided separately in section V in order to explain fully the fixed point structure found for arbitrary bulk viscosity exponents. 
In addition, section VI displays numerical evolution of some bulk viscosity models along with estimation of best-fit values of their free parameters using the Markov Chain Monte Carlo method. Finally, main findings and a general discussion of this work are presented in section VII.

\section{The bulk viscosity and its parametrizations}

Most of the approximations to describe a non-perfect fluid, that can serve to account for the unknown DM component, have implemented different phenomenological parameterizations for the bulk viscosity.  These parameterizations have covered from the simplest scenario, i.e., $\xi = \xi_0$, to more involved descriptions in terms of the Hubble rate and its time derivatives \cite{Ren:2005nw}. Accordingly, most of the  Ans\"{a}tze used in the literature take into account any of these three basic functional forms: i) a constant bulk viscosity coefficient, $\xi_0$, ii) a power of the (dark matter) energy density, and iii) a lineal function of the Hubble parameter and/or its times derivatives. Nevertheless, other parameterizations have considered even polynomial and hyperbolic functions of the redshift \cite{xin2009friedmann,hernandez2019cosmological}. A $\Lambda$CDM model with this kind of parameterization has been constrained in  \cite{herrera2020constraints}.

On the other hand, bulk viscosity is a concept closely connected with the equation of state (EoS) used to close the cosmological system of equations. Thus, it is quite reasonable to expect that bulk viscosity can be written proportional to (power of) the density of the full system or constituents of it. Even more, the kinetic inner state of a system also contribute to the definition of viscosity. Such modifications are encoded into the temperature and then the viscosity is, in general, written in terms of the density and/or the temperature \cite{librocaro}. Motivated by this physical argument, one of the most common ways to parameterize the bulk viscosity is $\xi=\xi_{0}\rho^{s}$, where $\xi_{0}>0$ is a bulk viscous constant. This particular type of parameterization has been widely investigated in \cite{Big.Bang,rho1,Brevik,Analysing,valorxi1,valorxi2,primerarticulo,sym14091866,Cruz:2022zxe}. Apart from these physical motivations, the power-law form allows to obtain analytical cosmological solutions. In particular, for the dissipative $\Lambda$CDM model with the specific election $s=1$, an interesting exact solution was obtained which asymptotically tends to a de Sitter expansion, in a certain region of the parameter space. Moreover it has not initial singularity, known as ``soft-Big Bang'' \cite{Softbang1,SoftBang2}.

Of course, in general, exact solutions can be obtained for some special values of $s$ and the qualitatively study of cosmological behavior, for arbitrary $s$, can be implemented using dynamical system analysis.  For example, studies of this type show that bulk-viscous inflation is possible in the truncated IS theory models \cite{coley1995qualitative}, and in the full IS theory a dynamical analysis was performed in  \cite{coley1996qualitative}. More complex cosmological scenarios where $\Lambda$ and $G$ were taken variables, in the causal framework, were analyzed in \cite{mak2002causal}.  A cosmological model that considered a universe filled with interacting DE and dissipative DM components, and radiation, was analyzed in the full IS formalism for the special case $s=1/2$ \cite{lepe2017dynamics}. Other study indicates that a  causal model of a universe filled with dissipative DM component shows accelerated phase for the case $s=1/2$, but the case with $s<1/2$ and $s>1/2$ are ruled out because they do not drive accelerated expansions \cite{mohan2020dynamical}. An interesting work analyzed a universe with viscous radiation and non-viscous dust in the framework of the nonlinear IS, for arbitrary $s$ \cite{acquaviva2015nonlinear}. 

The exploration in the context of the non-causal approach, has allowed to include more general expressions for the bulk viscosity such a dependence with the Hubble parameter or combination including terms like  $\ddot{a}/aH$. However, these two last cases have failed in displaying the conventional radiation dominated phase or a matter one \cite{sasidharan2015bulk,sasidharan2016phase}. A model with viscous DM, taking $s=1/2$, and perfect fluids for dark energy and radiation, which also included an interaction term between DM and DE, was analyzed by performing stability analysis \cite{hernandez2020stability}.  
Taking advantage of some of the aforementioned progresses in this viscous scenarios, we introduce a new and non-trivial parameterization of the bulk viscosity able to produce exact analytic solutions. To be more precise, we introduce an Ansatz which combines the dependence on both the dark matter density and the Hubble parameter simultaneously, i.e., 
$\xi \sim  H^{1-2s} \rho_m^s$. This non-trivial form leads to a direct coupling between dark matter and all other components through the bulk viscosity. It implies, by construction, that the bulk viscosity effect becomes effective only in cosmological stages when dark matter is dominant and its energy density is proportional to the bulk viscosity coefficient. The latter condition is largely dictated by the dynamical behavior of the system. Another advantage of this functional form is the possibility of studying, in a general way, the phenomenological implications for any $s$-exponent through dynamical system analysis which 
is not possible for most of phenomenological parametrizations used in the literature.

\section{The model}
The main ingredients of the model are described by an effective fluid picture, containing radiation, bulk viscosity dark matter with non-vanishing effective pressure and the cosmological constant. According to this cosmological setup, the Friedmann and the acceleration equations are respectively written as
\begin{eqnarray}
    3H^{2}&=&8\pi G_{N} \left(\rho_{r}+\rho_{m}+\rho_{\Lambda}\right),\\ 
    3H^{2}+2\dot{H}&=&-8\pi G_{N} \left(P_{r}+P_{m}^{\rm eff}+P_{\Lambda}\right).\label{sec2:eqn1}
\end{eqnarray}
Here the usual polytropic relations for radiation $P_{r}=\rho_{r}/3$ and for the cosmological constant, with energy density $\rho_{\Lambda} \equiv \frac{\Lambda}{8 \pi G_N}$
(where $G_N$ is the Newton's constant), 
$P_{\Lambda}=-\rho_{\Lambda}$ are set.  
Nevertheless, we assume for the dark matter fluid a bulk viscous pressure $\Pi$ that gives place to a minimal extension of the $\Lambda$CDM model: 
\begin{equation}
    P_{m}^{\rm eff}=P_{m}+\Pi=-3H\xi,\label{sec2:eqn2}
\end{equation}
where $\xi$ is the usual bulk viscosity coefficient that obeys the second law of thermodynamics provided that $\xi>0$. We want to realize a qualitative examination of the physical properties of  general power-law bulk viscous models demanding, besides, that the bulk viscosity effect is consistently handled so that it naturally turns
off when matter density vanishes. For instance, viscous models with $\xi\propto H$ suffer from this problem on its own. To avoid this, we propose the very convenient non-standard functional form to express the bulk viscosity coefficient $\xi$ 
\begin{equation}
    \xi \equiv \frac{\xi_{0}}{8\pi G_{N}} H^{1-2s} H_{0}^{2s}\left(\frac{\rho_{m}}{\rho_{m,0}} \right)^{s}=\frac{\hat{\xi}_{0}}{8\pi G_{N}} H\; \Omega_{m}^{s},\label{sec2:eqn3}
\end{equation}
which allows us to write, in turns, the system of equations Eqns.~(\ref{sec2:eqn1})-(\ref{sec2:eqn3}) in the form of autonomous system for any value of the exponent $s$ with the aid of the combination of the Hubble parameter and the matter energy density. One may think \textit{a priori} that the dependence of $\xi$ on the Hubble parameter leads in turn to an explicit dependence on the other components.  Nevertheless, the precise combination of $H$ and $\rho_{m}$ makes the bulk viscosity exist in an effective way only when matter energy density is dominant. it means that the bulk viscosity is effectively turned off when $\rho_{m}$ vanishes. For instance, in the radiation domination era: $\rho_{m}\to0$ which leads naturally to $\xi\to0$.  Hence,  the  contribution  of  the  other  components  to  the  bulk  viscosity is  unimportant  at  leading  order  in  the  cosmological  background  evolution.

Notice that both $\xi_{0}$ and $\hat{\xi}_{0}=\frac{\xi_{0}}{(\Omega_{m,0})^{s}}$, with $\Omega_{m,0} \equiv \frac{8\pi G_{N} \rho_{m,0}}{3H_{0}^{2}}$, are dimensionless parameters within this setup, describing the bulk viscosity effect and it is the only new free parameter that accounts for the extension of the $\Lambda$CDM model. This new parametrization provides the advantage of exploring unconventional values of $s$ as negative ones as shall be shown later. We remind that in the widely used parameterization $\xi\propto\rho^{s}$, the most studied cases $s=0$ and $s=1/2$ have been investigated in the framework of dynamical analysis, nevertheless for different $s$ values this analysis has not been carried out due to the difficulty in writing the resulting equations in the autonomous form.  It is interesting to see that well-known viscous models are enclosed within this new form and they are part of one-parameter family of viscous cosmological solutions as the dynamical system analysis will reveal. For instance, $s=0$ leads evidently to the viscous model $\xi\propto H$, while $s=1/2$ leads to the particular case of the widely used parametrization $\xi\propto\rho_m^{1/2}$, i.e. dependence on the energy density, $\rho_m$, only through a power law.

The conservation equations can be formulated in the simple form
\begin{align}
    \dot{\rho}_{r} + 4 H \rho_{r} =0,\label{sec2:eqn4.1} \\ 
 \dot{\rho}_{m} + 
  3 H (\rho_{m} + \Pi) =0.\label{sec2:eqn4}
\end{align}
In order to solve the system, $s$ must be certainly specified, but, at which level of difficulty the system can be solved for arbitrary large value of $s$? Are the corresponding fixed points stable? If so, how different are such solutions from the $\Lambda$CDM model and from each other at the background level? These are ones of the features we want to investigate in the present paper by studying the  stability properties of the fixed points by the standard linear stability theory in the next section.

\section{Dynamical system analysis}\label{sec:dynsys:symbolic}
We start by defining the dimensionless variables that set the phase space of the system and allows us to rewrite it in the form of an autonomous system. They are defined so that they correspond to the energy density parameters associated to each fluid 
\begin{align}
& \Omega_{r} \equiv \frac{8\pi G_{N} \rho_{r}}{3H^{2}}; \;\; \Omega_{m} \equiv \frac{8\pi G_{N} \rho_{m}}{3H^{2}}; \;\; \Omega_{\Lambda} \equiv \frac{8\pi G_{N} \rho_{\Lambda}}{3H^{2}}.
\label{sec3:eqn1}
\end{align}
So, the Friedmann constraint takes the usual form
\begin{equation}
\Omega_{r}+\Omega_{m}+\Omega_{\Lambda}=1.\label{sec3:eqn2}
\end{equation}
From the continuity equations Eqns.~(\ref{sec2:eqn4.1}) and (\ref{sec2:eqn4}), the evolution equations for radiation and dark matter are, respectively, derived with the help of the acceleration equation  Eqn.~(\ref{sec2:eqn1}) (or in its alternative form given by Eqn.~(\ref{sec3:eqn4}) defined below) that introduces an explicit dependence on bulk viscosity. This also affects the evolution of the energy density parameter associated with the cosmological constant Eqn.~(\ref{sec3:eqn1}) since it is normalized by the Hubble parameter\footnote{It does not mean however that the cosmological constant evolves itself since the condition $\dot{\rho}_{\Lambda}=0$ is preserved at any time.}. After some algebraic manipulations,  the dynamical system is described as follows
\begin{align}
\Omega_{r}^\prime&=\Omega_{r} (-1  - 3 \hat{\xi}_{0} \Omega_{m}^{s} + \Omega_{r} - 3 \Omega_{\Lambda}) , \nonumber \\
\Omega_{m}^\prime&= 3 \hat{\xi}_{0} \Omega_{m}^{s} - 3 \hat{\xi}_{0} \Omega_{m}^{1 + s} + \Omega_{m} (\Omega_{r} - 3 \Omega_{\Lambda}),   \label{sec3:eqn3}\\
\Omega_{\Lambda}^\prime&=\Omega_{\Lambda}\left( - 
 3 (-1 + \Omega_{\Lambda}+\hat{\xi}_{0} \Omega_{m}^{s})  + \Omega_{r}\right).\nonumber
\end{align}
In this form, it is evidenced how the bulk viscosity may affect non-trivially the dynamical behavior of all physical quantities\footnote{Notice however that all components are (minimally) coupled to gravity whereby the latter acts as a messenger between them. This is an indirect way where the bulk viscosity effects may be present in different cosmological stages.} (\ref{sec3:eqn1}). Here the prime denotes derivative with respect to $N\equiv\ln a$. In the limit of $\hat{\xi}_{0}\to0$ the $\Lambda$CDM model is recovered as can be plainly checked.
We should mention that the autonomous system \eqref{sec3:eqn3} clearly can be written as part corresponding to the $\Lambda$CDM model, plus the bulk viscosity sector, which extends the standard cosmological realization. From the functional form of the bulk viscosity, a non-linear ``interaction"-like term emerge naturally. 
%
Less evident is that the nonlinearity in the autonomous system affects each equation differently as a consequence of our Ansatz. 

Notice that the evolution equation for radiation is an auxiliary equation that can be taken away from the system by using the Friedmann constraint Eqn.~(\ref{sec3:eqn2}). So the system is reduced to two dimensional phase space.

The effective EoS parameter is defined as
\begin{equation}
w_{\rm eff}=-\frac{2}{3}\frac{H^\prime}{H}-1, \quad \text{with}\quad \frac{H^\prime}{H}=\frac{1}{2} (-3 + 3 \hat{\xi}_{0}  \Omega_{m}^{s} - \Omega_{r} + 3  \Omega_{\Lambda}),\label{sec3:eqn4}
\end{equation}
where the new term related to the bulk viscosity appears here explicitly for a general exponent $s$. One expects, from physical reasons, that $\hat{\xi}_{0}< 1$ (other than thermodynamics arguments $\hat{\xi}_{0}>0$), as has been also confirmed by different observational constraints. Nevertheless, one can adopt a less conservative position regarding the bulk viscosity magnitude what leads to another cosmological scenarios. Accordingly, we will refer to as the strong viscous regime when $\hat{\xi}_{0}\sim\mathcal{O}(1)$ which genuinely resembles unified dark fluid scenarios that can generate accelerated expansion on its own.  Without taking in advance any prejudice against its magnitude, we will see that these kind of solutions appear naturally in the model so that one could in principle abandon the accelerating mechanism behind traditional dark energy models as the one provided by the cosmological constant. Thought this not the main concern of this paper, the dynamical system analysis will allows us to study such solutions in a joint manner along with the proposed scenario of dissipative $\Lambda$CDM model. Since such viscous accelerating solutions correspond to a particular region of the parameter space they will be also discussed for the sake of completeness.

For practical purposes we have delimited the range of values for the exponent to $-2<s<2$ to be discussed in this section since there are \textit{a priori} no fundamental arguments to choose a different interval. It is worthwhile pointing out however that we have considered a larger range of $s$ for viscous models, thought not reported here, whose properties have been obtained in a systematic way as well by using symbolic programming in Wolfram Mathematica \cite{Mathematica}. Thus, for any range (or value) of $s$ the dynamical system can be solved by implementing the algorithm used.

\begin{figure*}
\centering
\includegraphics[width=0.32\hsize,clip]{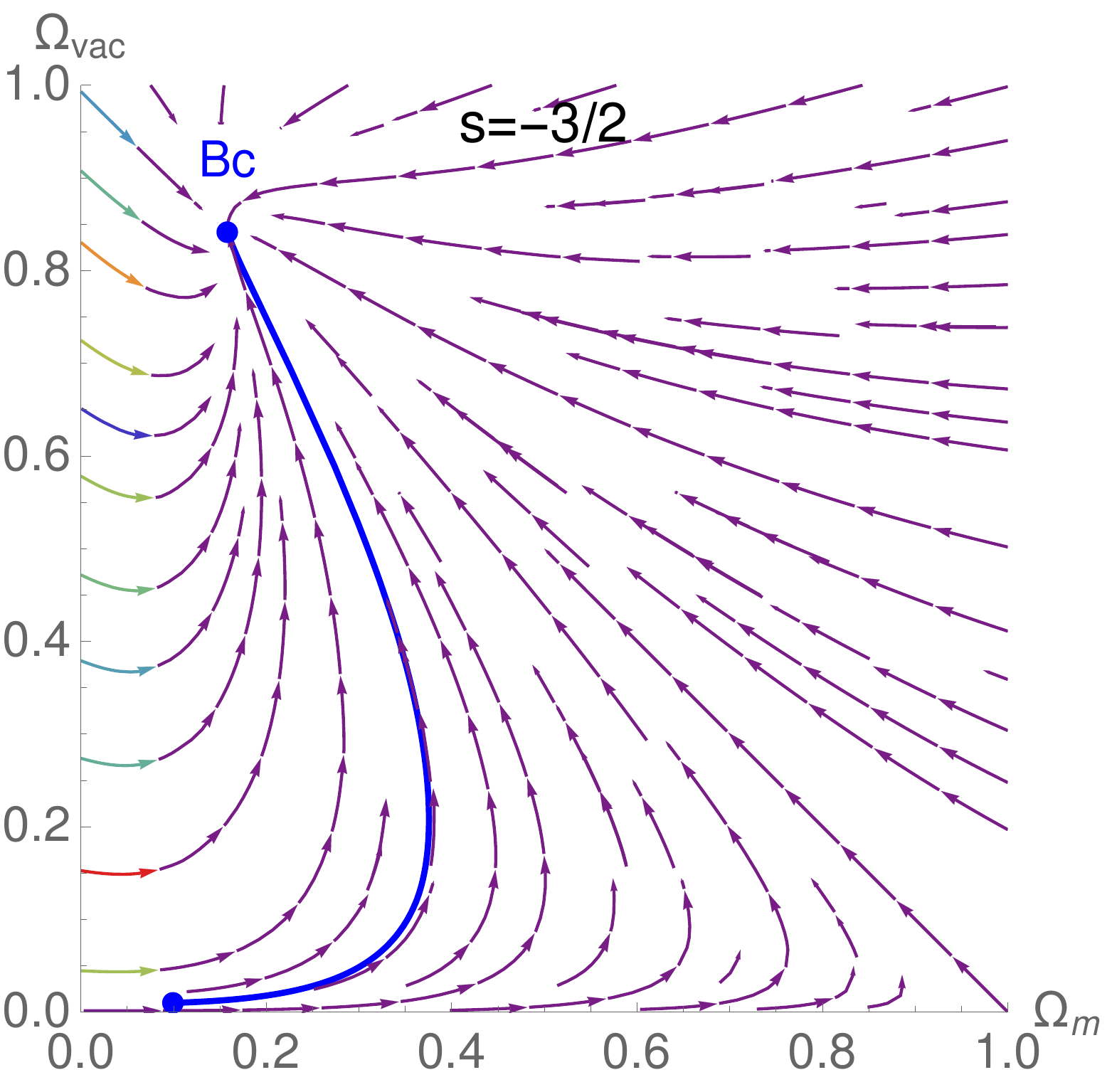}
\includegraphics[width=0.32\hsize,clip]{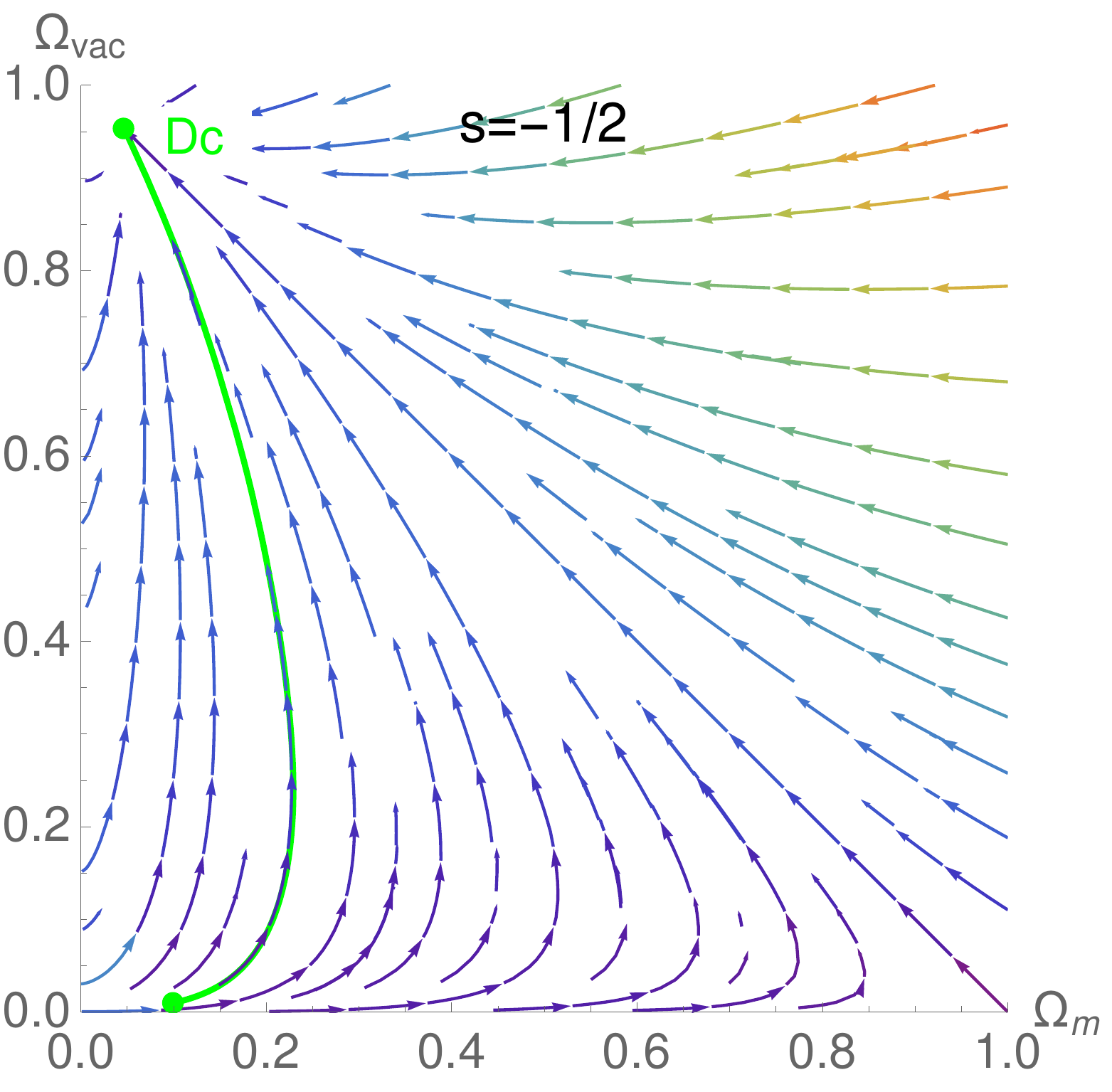}
\includegraphics[width=0.32\hsize,clip]{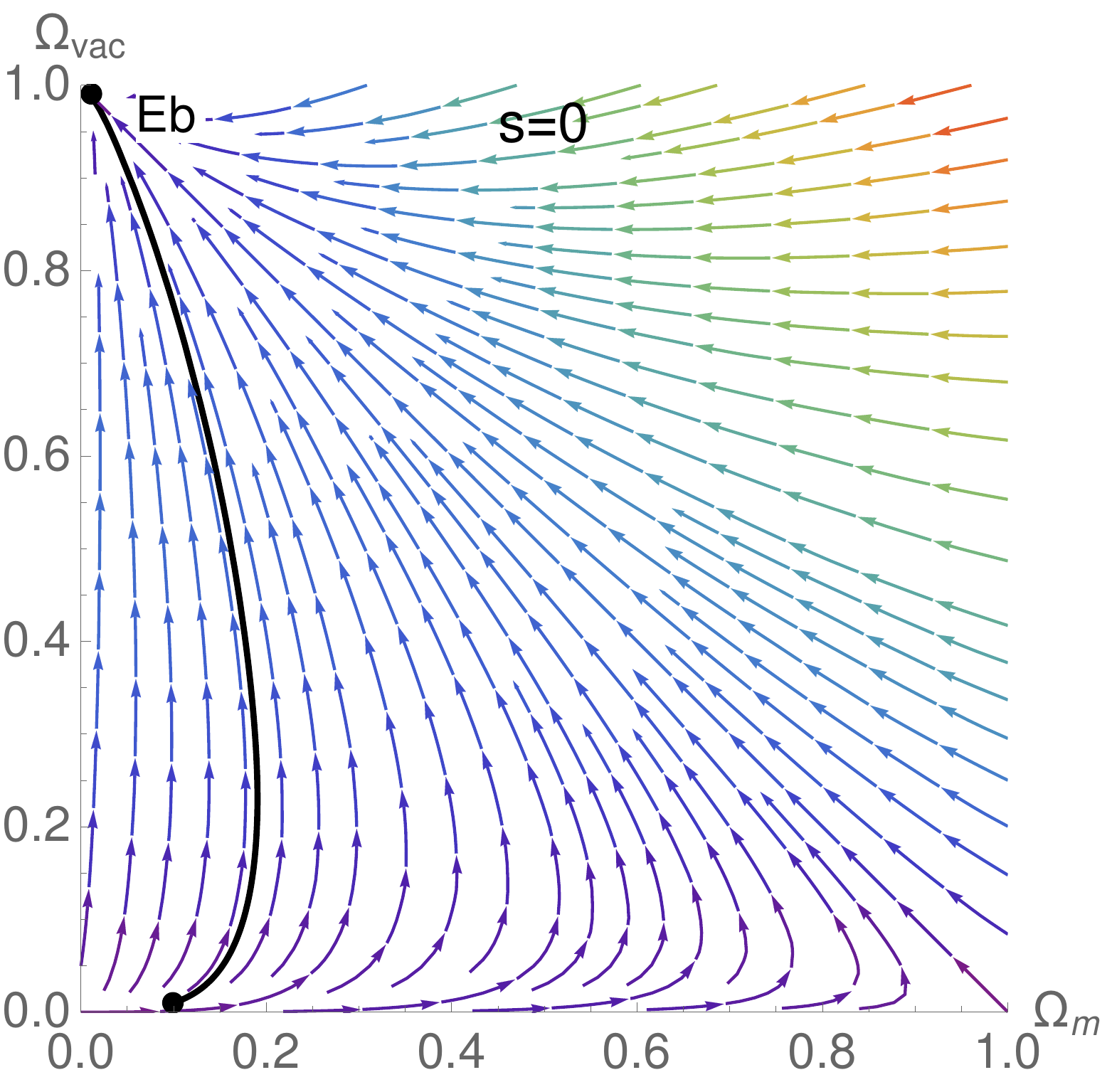}
\includegraphics[width=0.32\hsize,clip]{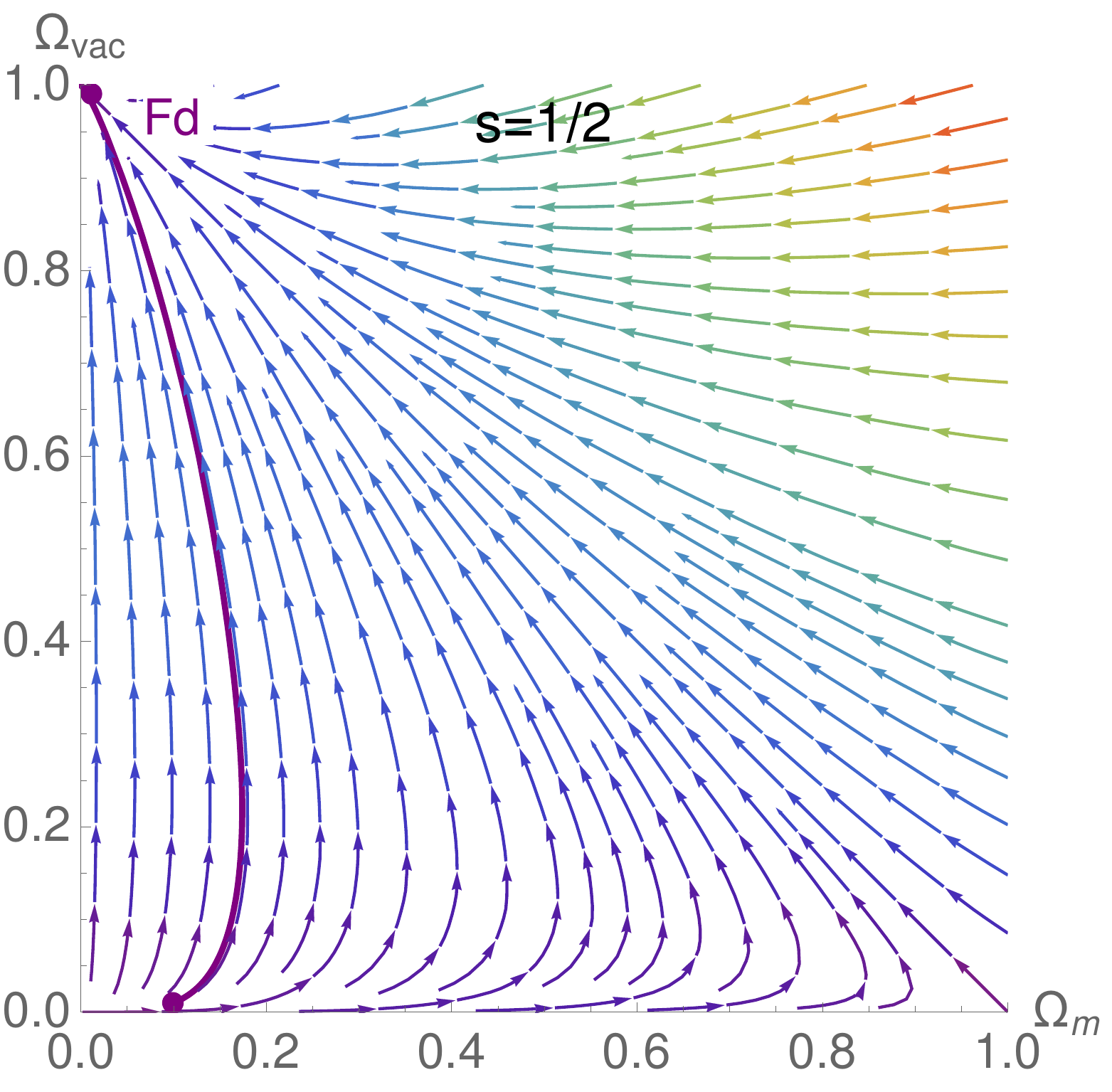}
\includegraphics[width=0.32\hsize,clip]{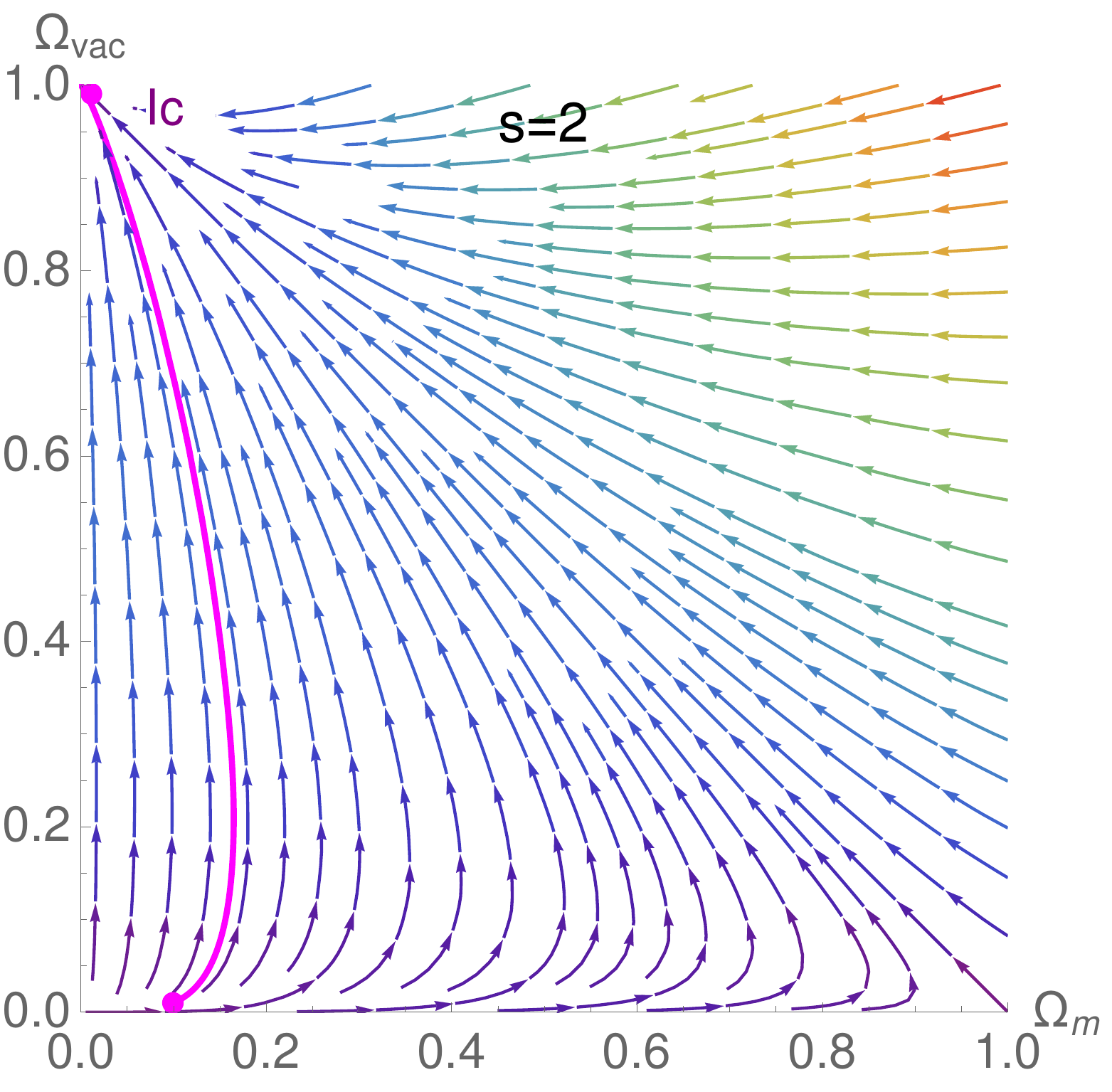}
\caption{Phase space diagrams of the system for $s=-3/2, -1/2, 0$ (top left, middle and right, respectively), $s=1/2, 2$ (bottom left and right, respectively) as indicated, along with their associated numerical trajectories, are presented. In all cases, we have chosen the initial conditions: $\Omega_{\rm m}^{(i)}= 0.1$, $\Omega_{\rm vac}^{(i)}= 0.01$, and $\hat{\xi}_{0}=0.01$. Additionally, we have displayed the corresponding fixed points Bc, Dc, Eb, Fd, Id, describing the de-Sitter solutions in accordance with Table \ref{M1:critical points}.} \label{fig:phase_space}
\end{figure*}

Fixed points for several viscous models with specific $s$-values are displayed in Table \ref{M1:critical points} along with their main cosmological features. A simple inspection points out that negative integers of $s$ are clearly discarded because they can not provide a complete cosmological dynamics: the radiation era, that is usually ignored for simplification reasons in most of the dynamical system studies, has no associated fixed point. So, the radiation era can not be unfortunately described by such viscous models because it is coupled directly to the viscous matter fields through the Hubble parameter in Eqn.~(\ref{sec2:eqn3}) other than gravity and, as a result, it prevents the solution to exist. A more precise mathematical reasoning will be addressed however in section V.  half-integer $s$-values, including the negatives ones, are, on the contrary, allowed with well defined (positive) energy densities parameters. 

The case $s=0$ is also inadmissible for the same fundamental reason as negative-integer values. This claim will be mathematically explained in the next section.
For a sufficiently large exponent values, including the sample shown here, the dynamical system analysis suffices to recognize the emergence of a first one-parameter family of viscous models, i.e., with a common fixed points structure, describing the current accelerated expansion with energy density parameters $\Omega_{\Lambda}=1-\hat{\xi}_{0}^{\frac{1}{1-s}}$ and $\Omega_{m}=\hat{\xi}_{0}^{\frac{1}{1-s}}$ with the restriction $s\neq1$. As this particular solution is not present for the case $s=1$, the system has to evolve for such a case towards fixed points with vanishing viscosity as the only suitable cosmological solution. This solution corresponds basically to the $\Lambda$CDM model with the possibility of having distinctively $w_{\rm eff}=-\hat{\xi}_{0}$ during  matter domination. This can be in principle troublesome unless one takes $\hat{\xi}_{0}$ small enough to have $w_{\rm eff}\sim 0$ or consider another physical interpretation. Interestingly, this solution can be identified as a unified dark fluid scenario that has the potential of driving the cosmic acceleration for large values of the bulk viscosity coefficient: $\hat{\xi}_{0}\sim\mathcal{O}(1)$. It is not very surprising to see that this kind of solution is intrinsically embedded in this model as one would expect. We do not explore, however, the cosmological implications of this solution since unified dark matter models are strongly constrained by different observations \cite{Benetti:2021div,2012PhLB..710...17L,2011JCAP...09..016V,2012JCAP...03..039C}. 
Although the derived conclusions from those works are not strictly applicable to the present model since we are taking a different parameterization for the bulk viscosity, we focus for the sake of concreteness on the viscous dark matter scenario as a minimal extension of the $\Lambda$CDM model.

Another common fixed point between these solutions appears during radiation domination and can be written generally as $\Omega_{r}=1-(3\hat{\xi}_{0})^{\frac{1}{1-s}}$ and $\Omega_{m}=(3\hat{\xi}_{0})^{\frac{1}{1-s}}$. The structure of this solutions prevent to take $s=1$, which implies that the bulk viscosity effect are absent during the radiation stage. Nevertheless, we will not respond at this point why for all negative integer $s$-values such a fixed point is not allowed and leave it once again to be tackled in section V. This peculiarity can be better appreciated in Table \ref{M1:critical points}. Another generality in this model is the existence of a twofold degeneracy in $w_{\rm eff}=\pm\hat{\xi}_{0}$ during matter domination for all half-integers $s$-values. This is indeed related to the possible real roots of a polynomial equation. This point will be better clarified in the next section in order to fully understand the pattern structure of the fixed points found. So, we have restricted ourselves to discuss the main cosmological implications of the fixed points.

As a general conclusion of the general structure of the fixed points, the bulk viscosity effect allows several cosmological trajectories to exist in phase space: ones of them following simply the standard evolution from radiation, matter and dark energy domination, and other ones allowing the existence of bulk viscosity during either all or some periods of the evolution of the universe. This general feature can also be appreciated in the phase space diagrams, shown in Fig.~\ref{fig:phase_space}, for various values of the $s$-exponent, encompassing both negative and positive values. These plots also illustrate the intriguing dynamical behavior of the systems we found in the asymptotic limit for $0<\hat{\xi}_{0}<1$: solutions with negative $s$-values lead to a non-vanishing $\Omega_{m}$ due to the bulk viscosity pressure, while for $s>1$, the bulk viscosity leads to large energy densities. We will extend this discussion when presenting the cosmological evolution for the considered viscous models.

Stability conditions can determine the available parameter space where the emerging fixed points can play a relevant role in the cosmological dynamics. The dynamical character of the fixed points can be checked by setting the (right) sign of the real parts of the eigenvalues associated to the Jacobian matrix of the the linear system. Following this criterion, the standard classifications of the fixed points are displayed in table \ref{M1:eigenvalues}. The condition $\hat{\xi}_{0}>0$ must be satisfied for all fixed points. Within this range, however, solutions whose eigenvalues are constrained by $0<\hat{\xi}_{0}<1$ belong to the viscous dark matter scenario while for the ample parameter space $\hat{\xi}_{0}>1$ the solutions are identified as unified dark fluid model. This is the reason why each fixed point can exhibit (up to) two distinct dynamical character. This is summarized in Table \ref{M1:eigenvalues}. We stress once again that the free parameter $\hat{\xi}_{0}$ for all dissipative models is barely constrained from dynamical system perspective. Nevertheless, this analysis provides a stringent restriction of what dissipative exponents $s$ are suitable for cosmological purposes.
\newpage

\begin{table*}[htp]
\centering  
\caption{Fixed points of the autonomous system described by Eqn.~(\ref{sec3:eqn3}) for different values of the bulk viscosity exponent $s$ along with the condition of existence of the fixed point. The main cosmological features of the model have been also included.}
\begin{ruledtabular}
\begin{tabular}{ccccccccccc}
Exponent& Point & $\Omega_{r}$ & $\Omega_{m}$ & $\Omega_{\Lambda}$ & $w_{\rm eff}$ & \text{Existence} & \text{Acceleration}
  \\ \hline
  \multirow{4}{*}{$s=-2$} &  
 $(\rm Aa)$ & $0$ & $1$ & $0$ & $-\hat{\xi}_{0}$ &  $\forall \hat{\xi}_{0}$ & $\text{Yes}$\\
 & $(\rm Ab)$ & $0$
 & $\hat{\xi}_{0}^{1/3}$ & $1-\hat{\xi}_{0}^{1/3}$ & $-1$ & $\forall  \hat{\xi}_{0}$ & $\text{Yes}$\\
 \\ \hline
  \multirow{4}{*}{$s=-\frac{3}{2}$} &  
 $(\rm Ba)$ & $0$ & $1$ & $0$ & $\hat{\xi}_{0}$ &  $\forall \hat{\xi}_{0}$ & $\text{No}$\\
 & $(\rm Bb)$ & $0$ & $1$ & $0$ & $-\hat{\xi}_{0}$ &  $\forall \hat{\xi}_{0}$ & $\text{Yes}$\\
 & $(\rm Bc)$ & $0$
 & $\hat{\xi}_{0}^{2/5}$ & $1-\hat{\xi}_{0}^{2/5}$ & $-1$ & $\forall \hat{\xi}_{0}$ & $\text{Yes}$\\
 & $(\rm Bd)$&$1-3^{2/5}\hat{\xi}_{0}^{2/5}$
 & $3^{2/5}\hat{\xi}_{0}^{2/5}$& $0$ & $\frac{1}{3}$ & $\forall \hat{\xi}_{0}$ & $\text{No}$\\
 \\ \hline
  \multirow{4}{*}{$s=-1$} &  
 $(\rm Ca)$ & $0$ & $1$ & $0$ & $-\hat{\xi}_{0}$ &  $\forall \hat{\xi}_{0}$ & $\text{Yes}$\\
 & $(\rm Cb)$ & $0$ & $\hat{\xi}_{0}^{1/2}$ & $1-\hat{\xi}_{0}^{1/2}$ & $-1$ &  $\hat{\xi}_{0}>0$ & $\text{Yes}$\\
 \\ \hline
  \multirow{4}{*}{$s=-\frac{1}{2}$} &  
 $(\rm Da)$ & $0$ & $1$ & $0$ & $\hat{\xi}_{0}$ &  $\forall \hat{\xi}_{0}$ & $\text{No}$\\
 & $(\rm Db)$ & $0$ & $1$ & $0$ & $-\hat{\xi}_{0}$ &  $\forall \hat{\xi}_{0}$ & $\text{Yes}$\\
 & $(\rm Dc)$ & $0$
 & $\hat{\xi}_{0}^{2/3}$ & $1-\hat{\xi}_{0}^{2/3}$ & $-1$ & $\hat{\xi}_{0}>0$ & $\text{Yes}$\\
 & $(\rm Dd)$ &$1-3^{2/3}\hat{\xi}_{0}^{2/3}$
 & $3^{2/3}\hat{\xi}_{0}^{2/3}$ & $0$ & $\frac{1}{3}$ & $\hat{\xi}_{0}>0$ & $\text{No}$\\
 \\ \hline
  \multirow{4}{*}{$s=0$} &  
 $(\rm Ea)$& $0$& $1$ & $0$ & $-\hat{\xi}_{0}$ & $\forall \hat{\xi}_{0}$ & $\text{Yes}$\\
 & $(\rm Eb)$& $0$
 & $\hat{\xi}_{0}$ & $1-\hat{\xi}_{0}$ & $-1$ & $\forall\hat{\xi}_{0}$ & $\text{Yes}$\\
 \\ \hline
  \multirow{4}{*}{$s=\frac{1}{2}$} &  
 $(\rm Fa)$ &$0$ & $1$ &$0$ & $\hat{\xi}_{0}$ & $\forall \hat{\xi}_{0}$ & $\text{No}$\\
 & $(\rm Fb)$  &$0$ & $1$ &$0$ &$-\hat{\xi}_{0}$ & $\forall \hat{\xi}_{0}$ & $\text{Yes}$ \\
 & $(\rm Fc)$ & $1-(3\hat{\xi}_{0})^{2}$
 & $(3\hat{\xi}_{0})^{2}$ & $0$ & $\frac{1}{3}$ & $\forall \hat{\xi}_{0}$ & $\text{No}$\\
 & $(\rm Fd)$ & $0$
 & $\hat{\xi}_{0}^{2}$& $1-\hat{\xi}_{0}^{2}$ & $-1$ & $\forall \hat{\xi}_{0}$ & $\text{Yes}$\\
 \\ \hline
  \multirow{4}{*}{$s=1$} &  
 $(\rm Ga)$ & $0$ & $1$ & $0$ & $-\hat{\xi}_{0}$ & $\forall \hat{\xi}_{0}$ & $\text{Yes}$\\
 & $(\rm Gb)$ & $1$
 & $0$ & $0$ & $\frac{1}{3}$ & $\forall \hat{\xi}_{0}$ & $\text{No}$\\
 & $(\rm Gc)$ & $0$
 & $0$ & $1$ & $-1$ & $\forall \hat{\xi}_{0}$ & $\text{Yes}$\\
 \\ \hline
  \multirow{6}{*}{$s=\frac{3}{2}$} &  
 $(\rm Ha)$ & $0$ & $1$ & $0$ & $\hat{\xi}_{0}$ & $\forall \hat{\xi}_{0}$ & $\text{No}$\\
 & $(\rm Hb)$ & $1$
 & $0$ & $0$ & $\frac{1}{3}$ & $ \forall \hat{\xi}_{0}$ & $\text{No}$\\
 & $(\rm Hc)$ & $0$
 & $0$ & $1$ & $-1$ & $\forall \hat{\xi}_{0}$ & $\text{Yes}$\\
 & $(\rm Hd)$ & $0$  & $1$ & $0$ & $-\hat{\xi}_{0}$ & $\forall \hat{\xi}_{0}$ & $\text{Yes}$\\
 & $(\rm He)$ & $1-(3\hat{\xi}_{0})^{-2}$
 & $(3\hat{\xi}_{0})^{-2}$ & $0$ & $\frac{1}{3}$ & $\hat{\xi}_{0}\neq0$ & $\text{No}$\\
 & $(\rm Hf)$ & $0$ & $\hat{\xi}_{0}^{-2}$ & $1-\hat{\xi}_{0}^{-2}$ & $-1$ & $\hat{\xi}_{0}\neq0$ & $\text{Yes}$\\
 \\ \hline
  \multirow{6}{*}{$s=2$} &  
 $(\rm Ia)$ & $0$ & $1$ & $0$ & $-\hat{\xi}_{0}$ & $\forall \hat{\xi}_{0}$ & $\text{Yes}$\\
 & $(\rm Ib)$ & $1$
 & $0$ & $0$ & $\frac{1}{3}$ & $\forall\hat{\xi}_{0}$ & $\text{No}$\\
 & $(\rm Ic)$ & $0$
 & $0$ & $1$ & $-1$ & $\forall \hat{\xi}_{0}$ & $\text{Yes}$\\
 & $(\rm Id)$ & $0$ & $\hat{\xi}_{0}^{-1}$ & $1-\hat{\xi}_{0}^{-1}$ & $-1$ & $\hat{\xi}_{0}\neq0$ & $\text{Yes}$\\
\end{tabular}
\end{ruledtabular}\label{M1:critical points}
\end{table*}
\begin{table*}[htp]
\centering  
\caption{Eigenvalues and stability conditions for setting the dynamical character of each fixed point corresponding to some bulk viscosity exponents $s$.}
\begin{ruledtabular}
\begin{tabular}{ccccccccccc}
Exponent & Point & $\lambda_{1}$ & $\lambda_{2}$ & \text{Stability} \\ \hline
 \multirow{2}{*}{$s=-2$} &
$(\rm Aa)$ & $-1-3\hat{\xi}_{0}$ & $3-3\hat{\xi}_{0}$ &$\text{Saddle}\;\text{if}\;0<\hat{\xi}_{0}<1; \text{atractor}\;\text{if}\;\hat{\xi}_{0}>1$\\
 & $(\rm Ab)$ & $-4$ & $9(-1+\hat{\xi}_{0}^{1/3})$ &$\text{Saddle} \;\text{if}\; \hat{\xi}_{0}> 1;\text{atractor}\;\text{if}\;0<\hat{\xi}_{0}<1$\\
\\ \hline
 \multirow{4}{*}{$s=-\frac{3}{2}$} &
$(\rm Ba)$ & $3(1+\hat{\xi}_{0})$ & $-1+3\hat{\xi}_{0}$ & $\text{Repeller}\;\text{if}\; \hat{\xi}_{0}> 1/3;\text{saddle} \;\text{if}\; 0<\hat{\xi}_{0}<1/3$\\
 & $(\rm Bb)$ & $-1-3\hat{\xi}_{0}$ & $3-3\hat{\xi}_{0}$ &$\text{Saddle}\;\text{if}\;0<\hat{\xi}_{0}<1; \text{atractor}\;\text{if}\;\hat{\xi}_{0}>1$ \\
& $(\rm Bc)$ & $-4$ & $\frac{15}{2} (-1 + \hat{\xi}_{0}^{2/5})$ &$\text{Saddle} \;\text{if}\; \hat{\xi}_{0}>1;\text{attractor} \;\text{if}\; 0<\hat{\xi}_{0}< 1$\\
& $(\rm Bd)$ & $4$ & $-\frac{5}{2} (-1 + (3 \hat{\xi}_{0})^{2/5})$ & $\text{Repeller} \;\text{if}\;0<\hat{\xi}_{0}<1/3;\text{saddle} \;\text{if}\;\hat{\xi}_{0}>1/3$\\
\\ \hline
 \multirow{4}{*}{$s=-1$} &
$(\rm Ca)$ & $-1-3\hat{\xi}_{0}$ & $3-3\hat{\xi}_{0}$ &$\text{Saddle}\;\text{if}\;0<\hat{\xi}_{0}<1; \text{atractor}\;\text{if}\;\hat{\xi}_{0}>1$ \\
 & $(\rm Cb)$&$-4$ & $6(-1 + \hat{\xi}_{0}^{1/2})$ & $\text{Saddle} \;\text{if}\; \hat{\xi}_{0}>1; \text{atractor}\;\text{if}\;0<\hat{\xi}_{0}<1$\\
\\ \hline
 \multirow{4}{*}{$s=-\frac{1}{2}$} &
$(\rm Da)$ & $3(1+\hat{\xi}_{0})$ & $-1+3\hat{\xi}_{0}$ & $\text{Repeller}\;\text{if}\; \hat{\xi}_{0}>1/3; \text{saddle}\;\text{if}\;0<\hat{\xi}_{0}<1/3$\\
 & $(\rm Db)$ & $-1 - 3 \hat{\xi}_{0}$ & $3 -3\hat{\xi}_{0}$ & $\text{Saddle} \;\text{if}\; 0<\hat{\xi}_{0}<1; \text{attractor}\;\text{if}\;\hat{\xi}_{0}>1$\\
& $(\rm Dc)$ & $-4$ & $\frac{9}{2} (-1 +\hat{\xi}_{0}^{2/3})$ &$\text{Saddle} \;\text{if}\; \hat{\xi}_{0}>1;\text{attractor} \;\text{if}\; 0<\hat{\xi}_{0}<1$\\
& $(\rm Dd)$ & $4$ & $-\frac{3}{2} (-1 + (3\hat{\xi}_{0})^{2/3})$ & $\text{Repeller} \;\text{if}\; 0<\hat{\xi}_{0}<1/3;\text{saddle} \;\text{if}\;\hat{\xi}_{0}>1/3$\\
\\ \hline
 \multirow{4}{*}{$s=0$} &
$(\rm Ea)$ & $-1-3\hat{\xi}_{0}$ & $3-3\hat{\xi}_{0}$ & $\text{Saddle} \;\text{if}\; 0<\hat{\xi}_{0}<1;\text{attractor} \;\text{if}\;\hat{\xi}_{0}>1$\\
 & $(\rm Eb)$ & $-4$ & $3 (-1 + \hat{\xi}_{0})$ & $\text{Saddle} \;\text{if}\; \hat{\xi}_{0}>1;\text{attractor} \;\text{if}\;0<\hat{\xi}_{0}<1$\\
 \\ \hline
 \multirow{4}{*}{$s=\frac{1}{2}$} &
$(\rm Fa)$ & $3(1+\hat{\xi}_{0}) $ & $-1+3\hat{\xi}_{0}$ & $\text{Repeller} \;\text{if}\; \hat{\xi}_{0}>1/3;\text{saddle} \;\text{if}\;0<\hat{\xi}_{0}<1/3$\\
 & $(\rm Fb)$ & $-1 - 3 \hat{\xi}_{0}$ & $3 -3 \hat{\xi}_{0}$ & $\text{Saddle} \;\text{if}\; 0<\hat{\xi}_{0}<1;\text{attractor} \;\text{if}\;\hat{\xi}_{0}>1$\\
& $(\rm Fc)$ & $4$ & $\frac{1}{2} (1 -(3 \hat{\xi}_{0})^{2})$ & $\text{Repeller} \;\text{if}\; 0<\hat{\xi}_{0}<1/3;\text{saddle} \;\text{if}\;\hat{\xi}_{0}>1/3$\\
& $(\rm Fd)$ & $- 4$ & $\frac{3}{2} (-1 +\hat{\xi}_{0}^{2})$ & $\text{Saddle} \;\text{if}\; \hat{\xi}_{0}>1;\text{attractor} \;\text{if}\;0<\hat{\xi}_{0}<1$\\
\\ \hline
 \multirow{4}{*}{$s=1$} &
$(\rm Ga)$ & $-1-3\hat{\xi}_{0}$ & $3-3\hat{\xi}_{0} $ & $\text{Saddle} \;\text{if}\; 0<\hat{\xi}_{0}<1;\text{attractor} \;\text{if}\;\hat{\xi}_{0}>1$\\
 & $(\rm Gb)$ & $4$ & $(1+3 \hat{\xi}_{0})$ & $\text{Reppeler} \;\text{if}\; \hat{\xi}_{0}>0$\\
& $(\rm Gc)$ & $-4$ & $ -3 (1 - \hat{\xi}_{0})$ & $\text{Saddle} \;\text{if}\; \hat{\xi}_{0}>1;\text{attractor} \;\text{if}\;0<\hat{\xi}_{0}<1$\\
\\ \hline
 \multirow{4}{*}{$s=\frac{3}{2}$} &
$(\rm Ha)$ & $3(1+\hat{\xi}_{0})$ & $-1+3\hat{\xi}_{0}$ & $\text{Repeller} \;\text{if}\; \hat{\xi}_{0}>1/3;\text{saddle} \;\text{if}\;0<\hat{\xi}_{0}<1/3$\\
 & $(\rm Hb)$ & $4$ & $1$ &$\text{Repeller} \; \forall\hat{\xi}_{0}>0$\\
& $(\rm Hc)$ & $-4$ & $-3 $ &$\text{Attractor} \; \forall\hat{\xi}_{0}>0$\\
& $(\rm Hd)$ & $-1-3\hat{\xi}_{0}$ & $3-3\hat{\xi}_{0}$ & $\text{Saddle} \;\text{if}\; 0<\hat{\xi}_{0}<1;\text{attractor} \;\text{if}\;\hat{\xi}_{0}>1$\\
& $(\rm He)$ & $4$ & $-\frac{3}{2}(1 -(3 \hat{\xi}_{0})^{-2})$ & $\text{Repeller} \;\text{if}\; 0<\hat{\xi}_{0}<1/3;\text{saddle} \;\text{if}\;\hat{\xi}_{0}>1/3$\\
& $(\rm Hf)$ & $-4$ & $\frac{3}{2}(1-(\hat{\xi}_{0})^{-2})$ & $\text{Saddle} \;\text{if}\; \hat{\xi}_{0}>1;\text{attractor} \;\text{if}\;0<\hat{\xi}_{0}<1$\\
\\ \hline
 \multirow{4}{*}{$s=2$} &
$(\rm Ia)$ & $-1-3\hat{\xi}_{0}$ & $3-3\hat{\xi}_{0}$ & $\text{Saddle} \;\text{if}\; 0<\hat{\xi}_{0}<1;\text{attractor} \;\text{if}\;\hat{\xi}_{0}>1$\\
 & $(\rm Ib)$ & $4$ & $\frac{1}{2}$ & $\text{Repeller} \; \forall\hat{\xi}_{0}>0$\\
& $(\rm Ic)$ & $-4$ & $-\frac{3}{2}$ & $\text{Attractor} \; \forall\hat{\xi}_{0}>0$\\
& $(\rm Id)$ & $-4$ & $3(1-\hat{\xi}_{0}^{-1})$ & $\text{Saddle} \;\text{if}\; \hat{\xi}_{0}>1;\text{attractor} \;\text{if}\;0<\hat{\xi}_{0}<1$\\
\end{tabular}
\end{ruledtabular}\label{M1:eigenvalues}
\end{table*}
%

%
%

\section{Analytical study of the fixed points structure \\ for arbitrary bulk viscosity exponents}\label{sec:dynsys:analytical}

Motivated by the pattern structure of the fixed points explained in the above section and summarized in Table \ref{M1:critical points} for different dissipation exponents, we will study analytically the fixed points of the differential equations (\ref{sec3:eqn3}), which govern the evolution of the system in the phase space variables. We will compute their stability properties as well, since they are relevant for the physical (cosmological) scenarios predicted by the model. 

We will compute the fixed points of the dynamical system (\ref{sec3:eqn3}), defined as its stationary points or the points in the phase space spanned by $(\Omega_{r},\Omega_{m},\Omega_{\Lambda})$ at which the right hand side (r.h.s.) expressions of these equations vanish. We further will apply a linear stability analysis to them taking into account the requirements of the general stability theorem by Malkin \cite{Malkin_52}, such that the conclusions remain valid beyond the linear analysis. To simplify the notation we introduce the variables $X = \Omega_{r}$, $Y = \Omega_{m}$ and $Z = \Omega_{\Lambda}$. The fixed points can be organized into three general classes as follows,

Type $I$:   

\begin{equation}
X_I = 0, \quad Y_I = 1, \quad Z_I = 0, \quad \text{with} \quad w_{eff} = - 1^s \hat{\xi}_{0},
\label{GFP_I}
\end{equation}

Type $II$:

\begin{align}
a)\quad X_{II}& = 0, \quad Y_{II} = \hat{\xi}_{0}^{ 1 / (1-s)}, \quad Z_{II} = 1- Y_{II} \quad \text{with} \quad w_{eff} = -1 \label{GFP_II}\\
b)\quad X_{II}& = 0, \quad Y_{II} = 0, \quad Z_{II} = 1 \quad \text{with} \quad w_{eff} = -1,
\label{GFP_II_b}
\end{align}

Type $III$:

\begin{align}
a)\quad X_{III}& = 1- Y_{III}, \quad Y_{III} = (-3 \hat{\xi}_{0})^{ 1 / (1-s)}, \quad Z_{III} = 0 \quad \text{with} \quad w_{eff} = 1/3 \label{GFP_III}\\
b)\quad X_{III}& = 1, \quad Y_{III} = 0, \quad Z_{III} = 0 \quad \text{with} \quad w_{eff} = 1/3.
\label{GFP_III_b}
\end{align} 
  
Now, we will illustrate how to obtain some fixed points displayed in Table \ref{M1:critical points} by using the general equations (\ref{GFP_I})-(\ref{GFP_III_b}), which describe all fixed points of the dynamical system. The additional advantage of this analytical deduction is that it allows explaining the origin of the degeneracy of the fixed points associated to positive and negative semi-integer $s$-values, which comes from the multivalued dependence of the effective EoS coefficient on the viscous exponent: $\omega_{eff} = -\hat{\xi}_0 Y^s + X / 3 - Z$.\\

i) $s = -2$. The fixed points turn out to be: 
\begin{align}
X_I&= 0, \quad Y_I = 1, \quad Z_I = 0, \quad \text{with} \quad w_{eff} = - \hat{\xi}_{0}, \\
X_{II}&= 0, \quad Y_{II} = \hat{\xi}_{0}, \quad Z_{II} = 1 - \hat{\xi}_{0}^{1/3}, \quad \text{with} \quad w_{eff} = -1. \label{I_II_s_minus_2}
\end{align}
which correspond to the points (Aa) and (Ab) respectively.  
Concerning the type $III$, the third fixed point should be given by (\ref{GFP_III}) 
\begin{equation}
X_{III} = 1 + (3 \hat{\xi}_{0})^{1/3}, \quad Y_{III} = -3\hat{\xi}_{0}, \quad Z_{III} = 0, \quad \text{with} \quad w_{eff} = 1 /3. 
\end{equation}
Nevertheless, this point must be discarded as by definition the phase variables $(X, Y, Z)$ must be real non-negative quantities, which is not fulfilled by the coordinate $Y$, provided $\hat{\xi}_{0} > 0$.
Finally, note that the cases b) of types $II$ and $III$  do not exist for this $s = -2$, as $(X, Y = 0, Z)$ does not correspond to a stationary point of the system for any arbitrary $X$ and $Z$ values.

ii) $s = -3/2$. Due to the existence of two real roots of $\sqrt1 = \pm 1$, there are two fixed points of type $I$ corresponding to the expression (\ref{GFP_I}), 
\begin{equation}
X_I= 0, \quad Y_I = 1, \quad Z_I = 0, \quad \text{with} \quad w_{eff} = \pm \hat{\xi}_{0}, 
\end{equation}
which are the fixed points (Ba) and (Bb).
There is only one real fixed point of type $II$ :
\begin{equation}
X_{II}= 0, \quad Y_{II} = \hat{\xi}_{0}^{2/5}, \quad Z_{II} = 1 - \hat{\xi}_{0}^{2/5}, \quad \text{with} \quad w_{eff} = -1,
\end{equation}
which corresponds to the point (Bc).

Finally, the type $III$ of fixed point (see Eqn. (\ref{GFP_III})) yields only one real value:
\begin{equation}
X_{III} = 1 - (3 \hat{\xi}_{0})^{2/5}, \quad Y_{III} = (3\hat{\xi}_{0})^{2/5}, \quad Z_{III} = 0, \quad \text{with} \quad w_{eff} = 1 /3. 
\end{equation}
This is the point (Bd) of Table I.

iii) $s = -1$. There is a first fixed point of type $I$ given by
\begin{equation}
X_I= 0, \quad Y_I = 1, \quad Z_I = 0, \quad \text{with} \quad w_{eff} = - \hat{\xi}_{0}, 
\end{equation}
which corresponds to the fixed point (Ca).
For the type $II$ fixed points, there is only one physical solution:
\begin{equation}
X_{II}= 0, \quad Y_{II} = \hat{\xi}_{0}^{1/2}, \quad Z_{II} = 1 - (\hat{\xi}_{0})^{1/2}, \quad \text{with} \quad w_{eff} = -1,
\end{equation}
which is the (Cb) fixed point.
Finally, for the type $III$ fixed point we obtain
\begin{equation}
X_{III} = 1 + \left(-3 \hat{\xi}_{0}\right)^{1/2}, \quad Y_{III} = (-3\hat{\xi}_{0})^{1/2}, \quad Z_{III} = 0, \quad \text{with} \quad w_{eff} = 1 /3, 
\end{equation}
which of course fails to be a fixed point as $X_{II}$ and $X_{II}$ are pure imaginary numbers.

iv) $s = 1/2$. The fixed point of type $I$ is twofold degenerated 
\begin{equation}
X_I= 0, \quad Y_I = 1, \quad Z_I = 0, \quad \text{with} \quad w_{eff} = \mp \hat{\xi}_{0}, 
\end{equation}
and corresponds to (Fa)-(Fb) of table I. 

The type $II$ corresponds to the (Fd) fixed point of Table I 
\begin{equation}
X_{II}= 0, \quad Y_{II} = \hat{\xi}_{0}^{2}, \quad Z_{II} = 1 - \hat{\xi}_{0}^{2}, \quad \text{with} \quad w_{eff} = -1.
\end{equation}

Finally, the type $III$ fixed point goes into (Fc) of Table I
\begin{equation}
X_{III} = 1 - (3 \hat{\xi}_{0})^{2}, \quad Y_{III} = -3\hat{\xi}_{0}, \quad Z_{III} = 0, \quad \text{with} \quad w_{eff} = 1 /3. 
\end{equation}

It is worthwhile pointing out that the apparent fixed points given by the Eqns.(\ref{GFP_II_b}) and (\ref{GFP_III_b}), which are predicted by the linear stability theory do not fulfill the requirements of Malkin's nonlinear stability theorem, and therefore they don't remain valid including higher orders. The condition on the functions $F_i(X, Y, Z)$ in a sufficiently small neighborhood of a fixed point $(X_*, Y_*, Z_*)$ for the validity of the stability analysis beyond the linear approximation is given by

\begin{equation}
\left|F_i (X, Y, Z) \right| \leq \mathcal{N} \left(X^{2} + Y^{2} + Z^{2}\right)^{1/2 + \alpha},
\end{equation}
where $\left|F_i \right|$ stands for the absolute value of the three functions appearing on the r.h.s of Eqn. (\ref{sec3:eqn3}), and $\mathcal{N}$ and $\alpha$ are positive constants. Clearly this condition is violated by $F_2$, and therefore the stationary point $Y_*=0$ is just an apparent fixed point of the system.

One alternative method, which allows to avoid this mathematical technicality, is to perform the variable change $\Tilde{Y} = Y^{2}$ and construct a new dynamical system. It turns out, that this system has automatically discarded the stationary point $\Tilde{Y} = 0$ as a component of a fixed point $(X, \Tilde{Y}, Z)$, and therefore, the $Y_* = 0$ is a spurious stationary point, which must be dismissed as component of a fixed point of the nonlinear dynamical system.

v) $s=1$. This particular case must be handle separately as the expressions of Eqns. (\ref{GFP_II}) and (\ref{GFP_III}) become singular. One finds the following three fixed points

\begin{equation}
X_I= 0, \quad Y_I = 1, \quad Z_I = 0, \quad \text{with} \quad w_{eff} = - \hat{\xi}_{0}, 
\end{equation}
which corresponds to (Ga) of Table I. 

\begin{equation}
X_{II}= 0, \quad Y_{II} = 1, \quad Z_{II} = 1, \quad \text{with} \quad w_{eff} = - \hat{\xi}_{0}, 
\end{equation}
which corresponds to (Gc) of Table I, and last, one finds the point (Gb) 

\begin{equation}
X_{III}= 1, \quad Y_{III} = 0, \quad Z_{III} = 0, \quad \text{with} \quad w_{eff} = 1/3. 
\end{equation}

Finally, we close this section formulating general conclusions on the fixed points of the autonomous dynamical system described by Eqn. (\ref{sec3:eqn3}) for arbitrary $s$-values. For non-positive integer exponents $s = -n $ with $n= 0,1,2,3 ...$ , the fixed points $(II)_b$ and $(III)_b$ do not exist, as $Y=0$ fails to be a stationary point of Eqn. (\ref{sec3:eqn3}). Moreover, as $1^{ 1 / (1+n)}$ has only two real roots ($\pm 1$) for $n$ odd, while it has only one root (1) for $n$ even, and $(-1)^{ 1 / (1+n)}$ has no real roots for $n$ odd and just one $(-1)$ for $n$ even: this follows from the fact that the $n+1$-roots of $-1$ built a regular polygon of $n+1$ sides on the unitary circle in the complex plane, and the vertices correspond exactly to these roots, with the first vertex located at the angle $\pi /(n+1)$. Hence, for $n$ odd no vertex coincides with the real axes, while for $n$ even, exactly one vertex hits the real axis at $-1$, which is not acceptable as $Y \geq 0$. One therefore conclude that the fixed point of Type $III$  doesn't fulfill the constraint of being non-negative real numbers and therefore must be discarded. This explains why the radiation-type of fixed points ($w_{eff}=1/3$) are missing in Table I for the non-positive integer values of $s$.

For a general negative irrational value $s$ holds that $1^{ 1 / (1 -s)}$  has infinite roots given by the expression $ \exp[2m \pi i / (1-s)] $ with $m \in \mathbb{Z}$. Nevertheless, from all of them there is only one real number at $m=0$ (1), whereas $(-1)^{ 1 / (1 -s)} = \exp[(2m+1) \pi i / (1-s)] $ with $m \in \mathbb{Z}$ has no real roots. We conclude that all of fixed points of Type $III$ a) must be discarded.

In summary, it is straightforward to find all fixed points of the dynamical system described by Eqn. (\ref{sec3:eqn3}) classified under Types $I$-$III$, and understand their degeneracy by using the general analytical expressions given by Eqns.  (\ref{GFP_I})-(\ref{GFP_III_b}) as it was made for the former cases. These analytical expressions for the fixed points have the main advantage of describing all the existing stationary points for an arbitrary given value of the bulk viscosity parameter $s$. This allows, besides, to obtain the physical interpretation of the fixed points as we will discuss later on. 

\subsection{Stability of the fixed points}
In this subsection we will study analytically the stability properties of some of the fixed points found in the former section, to illustrate the method. To this aim, we will use the linear stability analysis suited in a small neighborhood of the critical points of the dynamical system (\ref{sec3:eqn3}), and we will require to compute the eigenvalues of the system associated Jacobian matrix. First order derivatives of the functions appearing on the right hand side   

\begin{align}
F_1(X,Y,Z)&=-X (1 + 3 \hat{\xi}_{0} Y^{s} - X + 3 Z) , \\
F_2(X,Y,Z)&= 3 \hat{\xi}_{0} (1 - Y )Y^{s} + Y (X - 3 Z),  \\
F_3(X,Y,Z)&= Z ( X + 3(1-Z) - 3\hat{\xi}_{0} Y^{s} ).
\end{align}

are straightforward computed as:

\begin{align}
F_{1,X}&=-1 - 3 \hat{\xi}_{0} Y^{s} - 3Z + 2X , \quad  F_{2,X} = Y, \quad  F_{3,X} = Z \label{lin_deriv_1}\\
F_{1,Y}&= -3s \; \hat{\xi}_{0} \; X \;Y^{s-1} , \quad F_{2,Y} = X-3Z + 3\hat{\xi}_{0} (s\; Y^{s-1}-(1+s) Y^{s}), \quad F_{3,Y} = -3s \; \hat{\xi}_{0} Y^{s-1} Z \label{lin_deriv_2}\\
F_{1,Z}&=-3X, \quad  F_{2,Z}= -3Y, \quad  F_{3,Z} = X - 3\hat{\xi}_{0} Y^{s} +3(1-2Z) \label{lin_deriv_3},
\end{align}

where the comma indicates, as usual, derivative with respect to the variable. We have to compute the characteristic equation for the eigenvalues for each fixed point
\begin{equation}
0 = \det (\mu I - a_{ij}), \quad \text{where} \quad a_{ij} = F_{i,j} ,    
\end{equation}

and analyze them to decide whether they correspond to attractors, repellers, or saddle points. This represents an independent test to the stability properties of the fixed points associated to the original dynamical system quoted in Tables \ref{M1:critical points} and \ref{M1:eigenvalues}, which were obtained by using symbolic programming in the Mathematica software. 

i) For $s= -2$, there are two fixed points. Let us start with (Aa) first, it follows

\begin{align}
F_{1,X}&=-1 - 3 \hat{\xi}_{0}, \quad  F_{2,X} = 1, \quad  F_{3,X} = 0 \\
F_{1,Y}&= 0, \quad F_{2,Y} = -3\hat{\xi}_{0}, \quad F_{3,Y} = 0\\
F_{1,Z}&=0, \quad  F_{2,Z}= -3, \quad  F_{3,Z} = 3(1-\hat{\xi}_{0}).
\end{align}
This leads to the following eigenvalues of the characteristic equation for $\mu$
\begin{equation}
\mu_1 = -1-3 \hat{\xi}_{0}, \quad \mu_2 = -3\hat{\xi}_{0}, \quad \mu_3 = 3(1 - \hat{\xi}_{0}),  
\end{equation}
and therefore, we conclude that the fixed point (Aa) is an attractor for $\hat{\xi}_{0}>1$, and a saddle point if $0 < \hat{\xi}_{0} < 1$, in agreement with the first row of Table II.
For the fixed point (Ab) we obtain

\begin{align}
F_{1,X}&=-4 , \quad  F_{2,X} = \hat{\xi}_{0}, \quad  F_{3,X} = 1-\hat{\xi}_{0}^{1/3} \\
F_{1,Y}&= 0, \quad F_{2,Y} = 3(-3+2\hat{\xi}_{0}^{1/3}), \quad F_{3,Y} = 6 (1-\hat{\xi}_{0}^{1/3})\\
F_{1,Z}&=0, \quad  F_{2,Z}= -3\hat{\xi}_{0}^{1/3}, \quad  F_{3,Z} = -6(1-\hat{\xi}_{0}^{1/3}).
\end{align}
The characteristic polynomial $p(\mu)= (\mu + 4)(\mu^2 + 3 \mu (4-3 \hat{\xi}_{0}^{1/3}) + 27 (1-\hat{\xi}_{0}^{1/3}))$ has three roots, corresponding to the following eigenvalues:
\begin{equation}
\mu_1 = -4, \quad \mu_2 = -9(1 - \hat{\xi}_{0}^{1/3}), \quad  \text{and} \quad \mu_3 = -3,  
\end{equation}
which corresponds to an attractor for $0<\hat{\xi}_{0}<1$, whereas for $\hat{\xi}_{0}>1$, it represents a saddle point. This conclusion agrees with row two of Table II.

ii) For $s=1/2$, as it was already mentioned in the above section, we perform a variable change $Y = \Omega^{1/2}$ and compute in a similar form the Jacobian of the resulting dynamical system, obtaining the eigenvalues of the characteristic equation corresponding to each fixed point. In fact, for the points (Fa)-(Fb) of Table I one obtains respectively

\begin{equation}
\mu_1 = -(1 \pm 3 \hat{\xi}_{0}), \quad \mu_2 = \mp 3\hat{\xi}_{0}, \quad \mu_3 = 3(1 \mp \hat{\xi}_{0}).  
\end{equation}
We therefore conclude that the fixed point (Fa) is an attractor in the two first directions, but a repeller in the $\mu_3$ direction, for $\xi < 1/3$. This conclusion agrees with the one of Table II, while (Fb) is a saddle point with two repeller directions $\mu_2$ and $\mu_3$ and $\mu_1$ being an attractor, provided $\xi_{0}< 1/3$. 

For the fixed point point (Fd) of Table I, we find the following eingenvalues
\begin{equation}
\mu_1 = -4, \quad \mu_2 = - \frac{3}{2} (1 - \hat{\xi}_{0}^2), \quad \mu_3 = -3,  
\end{equation}
which describes an attractor in all three directions, and finally, for (Fc) one finds a repeller, provided $\xi_{0} < 1/3$
\begin{equation}
\mu_1 = 4, \quad \quad \mu_2 = 1, \quad  \mu_3 = \frac{1}{2} (1 - 9\hat{\xi}_{0}^2). 
\end{equation}

iii) Other physically interesting fixed points arise for $s=1$, whose stability properties we will study in what follows. As for this particular $s$-value the dynamical system is described by Eqn. (\ref{sec3:eqn3}), it yields

\begin{align}
F_{1,X}&=-1 - 3 \hat{\xi}_{0} Y - 3Z + 2X , \quad  F_{2,X} = Y, \quad  F_{3,X} = Z \\
F_{1,Y}&= -3\hat{\xi}_{0} X , \quad F_{2,Y} = (X-3Z) + 3\hat{\xi}_{0} (1-2Y), \quad F_{3,Y} = -3 \hat{\xi}_{0} Z\\
F_{1,Z}&=-3X, \quad  F_{2,Z}= -3 Y, \quad  F_{3,Z} = X - 3\hat{\xi}_{0} Y+ 3(1 - 2 Z).
\end{align}

For the first fixed point (Ga): $X=0$, $Y=1$, and $Z=0$, the eigenvalues are
\begin{equation}
\mu_1 = -(1+3\hat{\xi}_{0}), \quad \mu_2 = -3\hat{\xi}_{0}, \quad  \text{and} \quad \mu_3 = 3(1-\hat{\xi}_{0}).  
\end{equation}
Thus, this fixed point corresponds to a saddle point for $0<\hat{\xi}_{0}<1$, and for $\hat{\xi}_{0}>1$ is an attractor. This conclusion agrees with the one of Table II.

For the fixed point (Gb), $X=1$, $Y=0$, and $Z=0$, the eigenvalues turn out to be
\begin{equation}
\mu_1 = 1, \quad \mu_2 = (1+3\hat{\xi}_{0}), \quad  \text{and} \quad \mu_3 = 4.  
\end{equation}
These eigenvalues indeed represent a repeller for $\hat{\xi}_{0}>0$, as claimed in Table II.

Finally, for the fixed point (Gc), $X=0$, $Y=0$, and $Z=1$, the eigenvalues are
\begin{equation}
\mu_1 = -4, \quad \mu_2 = -3(1-\hat{\xi}_{0}), \quad  \text{and} \quad \mu_3 = -3.  
\end{equation}
These eigenvalues represent an attractor for $0<\hat{\xi}_{0}<1$, and for $\hat{\xi}_{0}>1$ it is a generalized saddle point behaving as a repeller in the  $Y$ direction.

Hence, this independent analytical treatment has served as a crosscheck of the derived results of section \ref{sec:dynsys:symbolic}. The present analysis can straightforwardly be extended to other values beyond the ones displayed in Table I.

\section{\label{Constraints}Cosmological Constraints}\label{sec:constraints}
In this section, we shall constrain the free parameters of the viscous model discussed in the above sections, in the context of dynamical system, with the Supernovae Ia (SNe Ia) data and observational Hubble parameter data (OHD). For the purpose of this paper suffices to include data at low redshift in order to quantify the bulk viscous effects given by the parametrization Eqn. (\ref{sec2:eqn3}). 

For the constraint, we compute the best-fit values of the free parameters of the model and their corresponding uncertainties with the \textit{affine-invariant Markov Chain Monte Carlo} method (MCMC) \cite{Goodman2010}, implemented in the pure-\textit{Python} code \textit{emcee} \cite{Foreman_Mackey_2013}, by setting 30 chains or ``walkers". As a convergence test, we compute at every $50$ steps the autocorrelation time $\tau_{\rm corr}$ of the chains, provided by the \textit{emcee} module. If the current step is greater than $50\tau_{\rm corr}$ and if the value of $\tau_{\rm corr}$ changes by less than $1\%$ with respect to its previous value, then we will consider that the chains have converged and the code will be stopped. We discard the first $5\tau_{\rm corr}$ steps as ``burn-in" steps and we flatten the chains, thin by about $\tau_{\rm corr}/2$. Additionally, we compute the mean acceptance fraction, whose value must be between 0.2 and 0.5 \cite{Foreman_Mackey_2013}, which can be modified by the stretch move, both ones provided by the \textit{emcee} module. It is important to mention that, for this Bayesian statistical analysis, we need to construct the following Gaussian likelihood:
\begin{equation}\label{likelihood}
    \mathcal{L}\propto\exp\left(-\chi_{i}^{2}/2\right),
\end{equation}
where $\chi_{i}^{2}$ is known as merit function and $i$ stands for each data set considered in the fit, namely, SNe Ia, OHD, and for the joint analysis SNe Ia+OHD.

For the OHD, we make use of the compilation provided by Magaña \textit{et al.} \cite{magana2018}, which consists of 51 Hubble data points in the redshift range $0.07\leq z\leq 2.36$. Accordingly, the merit function for this data is constructed as
\begin{equation}\label{OHDmerit}
    \chi_{\rm OHD}^{2}=\sum_{i=1}^{51}{\left[\frac{H_{i}-H_{th}(z_{i},\theta)}{\sigma_{H,i}}\right]^{2}},
\end{equation}
where $H_{i}$ is the observational Hubble parameter at redshift $z_{i}$ with an associated error $\sigma_{H,i}$, provided by the OHD sample, $H_{th}$ is the theoretical Hubble parameter at the same redshift, and $\theta$ accounts for the  free parameters of the Viscous model. In this regard, the current value of the Hubble parameter $H_{0}$ is considered as an additional free parameter, which is written as $H_{0}=100\frac{km/s}{Mpc}h$ in terms of the dimensionless parameter $h$. For the latter, we consider the Gaussian prior given by $h\in G(0.7403,0.0142)$, according to the value obtained for $H_{0}$ by A. G. Riess \textit{et al.} \cite{Riess:2019cxk}, measured with a $1.91\%$ of uncertainty in a model independent way.

On the other hand, for the SNe Ia data, we make use of the Pantheon sample \cite{Pan-STARRS1:2017jku}, which consists of 1048 data points in the redshift range $0.01\leq z\leq 2.3$. Accordingly, the merit function for this data is constructed as
\begin{equation}\label{SNemerit}
    \chi_{\rm SNe}^{2}=\sum_{i=1}^{1048}\left[\frac{\mu_{i}-\mu_{th}(z_{i},\theta)}{\sigma_{\mu,i}} \right]^{2},
\end{equation}
where $\mu_{i}$ is the observational distance modulus of each SNe Ia at redshift $z_{i}$ with an associated error $\sigma_{\mu.i}$, $\mu_{th}$ is the theoretical distance modulus at the same redshift, and $\theta$ encompasses the free parameters of the Viscous model. The theoretical distance modulus for a spatially flat FLRW space-time takes the form
\begin{equation}\label{muth}
    \mu_{th}(z_{i},\theta)=5\log_{10}{\left[\frac{d_{L}(z_{i},\theta)}{Mpc}\right]+\Bar{\mu}},
\end{equation}
where $\Bar{\mu}=5\left[\log_{10}{(c)}+5\right]$, with $c$ the speed of light given in units of $km/s$, and $d_{L}$ is the luminosity distance given by
\begin{equation}\label{luminosity}
    d_{L}(z_{i},\theta)=(1+z_{i})\int_{0}^{z_{i}}{\frac{dz'}{H(z',\theta)}}.
\end{equation}

In the Pantheon sample, the distance estimator is obtained using a modified version of the Tripp's formula \cite{Tripp:1997wt}, with two nuisance parameters calibrated to zero with the method ``BEAMS with Bias Correction'' (BBC) proposed by Kessler and Scolnic \cite{Kessler:2016uwi}. Hence, the observational distance modulus is given by the expression
\begin{equation}\label{muobs}
    \mu_{i}=m_{B,i}-\mathcal{M},
\end{equation}
where $m_{B,i}$ is the corrected apparent B-band magnitude of a fiducial SNe Ia at redshift $z_{i}$, provided by the Pantheon sample, and $\mathcal{M}$ is a remaining nuisance parameter which must be jointly estimated with the free parameters $\theta$ of the Viscous model. Therefore, the merit function for the SNe Ia data given by Eqn. \eqref{SNemerit} can be rewritten, in matrix notation (denoted by bold symbols), as
\begin{equation}\label{matrixSNe}
    \chi_{\rm SNe}^{2}=\textbf{M}(z,\theta,\mathcal{M})^{\dagger}\textbf{C}^{-1}\textbf{M}(z,\theta,\mathcal{M}),
\end{equation}
where $[\textbf{M}(z,\theta,\mathcal{M})]_{i}=m_{B,i}-\mu_{th}(z_{i},\theta)-\mathcal{M}$ and $\textbf{C}=\textbf{D}_{\rm stat}+\textbf{C}_{\rm sys}$ is the total uncertainties covariance matrix, with $\textbf{D}_{\rm stat}=diag(\sigma_{m_{B,i}}^{2})$ is the statistical uncertainties of $m_{B}$ and $\textbf{C}_{\rm sys}$ is the systematic uncertainties in the BBC approach, provided by the Pantheon sample.

Finally, in order to reduce the number of constants and free parameters, we marginalize over the nuisance parameter $\bar{\mathcal{M}}=\bar{\mu}+\mathcal{M}$, expanding the merit function \eqref{matrixSNe} as \cite{Lazkoz:2005sp}
\begin{equation}\label{ExpandedSNe}
    \chi^{2}_{\rm SNe}=A(z,\theta)-2B(z,\theta)\bar{\mathcal{M}}+C\bar{\mathcal{M}}^{2},
\end{equation}
where
\begin{equation}\label{defofA}
    A(z,\theta)=\textbf{M}(z,\theta,\bar{\mathcal{M}}=0)^{\dagger}\textbf{C}^{-1}\textbf{M}(z,\theta,\bar{\mathcal{M}}=0),
\end{equation}
\begin{equation}\label{defofB}
    B(z,\theta)=\textbf{M}(z,\theta,\bar{\mathcal{M}}=0)^{\dagger}\textbf{C}^{-1}\textbf{1},
\end{equation}
\begin{equation}\label{defofC}
    C=\textbf{1}\textbf{C}^{-1}\textbf{1}.
\end{equation}
Hence, by minimizing the expanded merit function \eqref{ExpandedSNe} with respect to $\bar{\mathcal{M}}$, gives $\bar{\mathcal{M}}=B(z,\theta)/C$, value that reduces the expanded merit function to
\begin{equation}\label{ReducedSNe}
    \chi^{2}_{\rm SNe}=A(z,\theta)-\frac{B(z,\theta)^{2}}{C},
\end{equation}
which corresponds to the merit function for the SNe Ia data used in our MCMM analysis that clearly depends only on the free parameters $\theta$ of the Viscous model.

It is important to mention that the merit function for the SNe Ia data given by the Eqn. \eqref{matrixSNe} provides the same information as the expanded and minimized merit function given by Eqn. \eqref{ReducedSNe}, since the best-fit values for the free parameters minimize the respective merit function. Therefore, the merit function evaluated in the values of the best-fit parameters, $\chi_{\rm min}^{2}$, give us a goodness-of-fit indicator independently of the data set used: the smaller the value of $\chi_{\rm min}^{2}$ is, the better the fit is. Nevertheless, one can in principle minimize the merit function by adding more free parameters to the model, resulting in over-fitting. In this sense, we can compare the goodness-of-fit in an statistical way by using the Bayesian criterion information (BIC) \cite{1978AnSta...6..461S}, which introduce a penalization on the value of $\chi^{2}_{\rm min}$ according to the expression
\begin{equation}\label{BIC}
    BIC=\chi^{2}_{\rm min}+\theta_{N}\ln{(n)},
\end{equation}
where $\theta_{N}$ is the number of free parameters of the Viscous model and $n$ is the total number of data points in the data sample used. Therefore, the model most favored statistically by observations, as compared to other, corresponds to the one with the smallest value of BIC, where a difference in $2-6$ in BIC between the two models is considered as evidence against the model with higher BIC, a difference of $6-10$ is already strong evidence, and a difference $>10$ correspond to a very strong evidence.

Since in the two data set the respective merit function depends mainly on the Hubble parameter as a function of the redshift (see Eqns. \eqref{OHDmerit} and \eqref{SNemerit}), then for the fit, we need to numerically integrate the Eqn. \eqref{sec3:eqn3} and \eqref{sec3:eqn4}. But, considering that the highest redshift in the combined OHD and the Pantheon samples is $z=2.36$, then the cosmological constraint is confined only to late-times where the radiation component is negligible ($\Omega_{r}\approx 0$) in comparison with the other components. The exclusion of $\Omega_{r}$ affects far beyond the percent level the best-fit values and make, in addition, the analysis more efficient in terms of time performance. Therefore, by using Eqn. \eqref{sec3:eqn2} and considering that $dN=-dz/(1+z)$, the system of differential equations reduces to
\begin{eqnarray}\label{ConstrainModel}
    \dfrac{dH}{dz} &=& \frac{3H}{2(1+z)}\Omega_{m}\left(1-\hat{\xi}_{0}\Omega_{m}^{s-1}\right), \\
    \dfrac{d\Omega_{m}}{dz} &=& \frac{3\Omega_{m}}{\left(1+z\right)}\left(1-\Omega_{m}\right)\left(1-\hat{\xi}_{0}\Omega_{m}^{s-1}\right), \nonumber
\end{eqnarray}
where $\Omega_{\Lambda}=1-\Omega_{m}$, and using as initial conditions $H(z=0)=H_{0}=100\frac{km/s}{Mpc}h$ and $\Omega_{m}(z=0)=\Omega_{m,0}$. Even more, for a further comparison, we also compute the best-fit parameters for the $\Lambda$CDM model, whose respective Hubble parameter as a function of the redshift at late-times is given simply by
\begin{equation}\label{LCDM}
    H(z) = 
    H_0
    \sqrt{\Omega_{m,0}(1+z)^{3}+1-\Omega_{m,0}}.
\end{equation}

The free parameters of the viscous model are $\theta=\{h,\Omega_{m,0},\hat{\xi}_{0},s\}$, while for the $\Lambda$CDM model are $\theta=\{h,\Omega_{m,0}\}$, for which, beside the Gaussian prior on $h$, we consider flat priors on the free parameters $\Omega_{m,0}$ and $\hat{\xi}_{0}$ given by $\Omega_{m,0}\in F(0,1)$ and $\hat{\xi}_{0}\in F(0,1)$, respectively. The latter range is mainly motivated by the dynamical system analysis described in Section \ref{sec:dynsys:symbolic}, which encloses the physical scenario we are interested in as an extension of the $\Lambda$CDM model. In this sense, we also consider the same fixed values for the free parameter $s$ used in the dynamical system analysis that provide a successful description of the complete cosmological dynamics. These are: $s=-3/2$, $-1/2$, $1/2$, $1$, $3/2$, and $2$. Hence, the cosmological constraint performed here serves as a complementary test to assess the cosmological viability of this class of viscous models.

In Table \ref{tab:MCMCParameters}, we present the total number of steps, the mean acceptance fraction (MAF), and the autocorrelation time $\tau_{\rm corr}$ for each free parameter model, obtained when the convergence test described in this section is fulfilled in the MCMC analysis for both, the $\Lambda$CDM model and the viscous model with  $s=-3/2$, $-1/2$, $1/2$, $1$, $3/2$, and $2$. It is important to mention that the values of the MAF are obtained for a value of the stretch move of $a=7$ for the $\Lambda$CDM model and $a=4.5$ for the viscous model, except for $s=-3/2$ where the value of the stretch moves to $a=3$.

\begin{table*}
    \centering
    \caption{\label{tab:MCMCParameters} Total number of steps, mean acceptance fraction (MAF), and autocorrelation time $\tau_{\rm corr}$ for each free parameter model, obtained when the convergence test described in the Section \ref{Constraints} is fulfilled in the MCMC analysis for the $\Lambda$CDM model and the Vicous model with $s=-3/2$, $-1/2$, $1/2$, $1$, $3/2$, and $2$; by setting $30$ chains or ``walkers'' for the SNe Ia data, OHD, and in their joint analysis. The values of the MAF are obtained for a value of the stretch move of $a=7$ for the $\Lambda$CDM model and $a=4.5$ for the Viscous model, except for $s=-3/2$ with $a=3$.}
    \begin{ruledtabular}
    \begin{tabular}{cccccc}
        \multirow{2}{*}{Data} & \multirow{2}{*}{Total steps} & \multirow{2}{*}{MAF} & \multicolumn{3}{c}{$\tau_{\rm corr}$} \\
        \cline{4-6}
         &  &  & $h$ & $\Omega_{m,0}$ & $\hat{\xi}_{0}$ \\
        \hline
        \multicolumn{6}{c}{$\Lambda$CDM model} \\
        SNe Ia & $1050$ & $0.365$ & $16.8$ & $17.4$ & $\cdots$ \\
        OHD & $1050$ & $0.364$ & $16.1$ & $16.2$ & $\cdots$ \\
        SNe Ia+OHD & $800$ & $0.365$ & $15.3$ & $15.1$ & $\cdots$ \\
        \hline
        \multicolumn{6}{c}{Viscous model $s=-3/2$} \\
        SNe Ia & $4900$ & $0.371$ & $41.4$ & $97.5$ & $91.0$ \\
        OHD & $5100$ & $0.388$ & $44.0$ & $98.3$ & $94.3$ \\
        SNe Ia+OHD & $5450$ & $0.405$ & $36.2$ & $108.0$ & $96.2$ \\
        \hline
        \multicolumn{6}{c}{Viscous model $s=-1/2$} \\
        SNe Ia & $2100$ & $0.342$ & $29.8$ & $40.0$ & $39.6$ \\
        OHD & $2300$ & $0.352$ & $32.6$ & $35.8$ & $35.6$ \\
        SNe Ia+OHD & $1650$ & $0.365$ & $27.0$ & $30.2$ & $29.0$ \\
        \hline
        \multicolumn{6}{c}{Viscous model $s=1/2$} \\
        SNe Ia & $2250$ & $0.344$ & $29.2$ & $42.3$ & $43.3$ \\
        OHD & $1950$ & $0.349$ & $28.1$ & $30.4$ & $33.7$ \\
        SNe Ia+OHD & $1550$ & $0.366$ & $24.6$ & $30.3$ & $29.6$ \\
        \hline
        \multicolumn{6}{c}{Viscous model $s=1$} \\
        SNe Ia & $2200$ & $0.338$ & $31.2$ & $43.7$ & $43.3$ \\
        OHD & $2200$ & $0.353$ & $33.7$ & $38.0$ & $39.1$ \\
        SNe Ia+OHD & $1700$ & $0.365$ & $28.8$ & $28.8$ & $28.2$ \\
        \hline
        \multicolumn{6}{c}{Viscous model $s=3/2$} \\
        SNe Ia & $2450$ & $0.333$ & $30.5$ & $47.9$ & $47.7$ \\
        OHD & $1750$ & $0.354$ & $31.8$ & $32.0$ & $33.0$ \\
        SNe Ia+OHD & $1600$ & $0.365$ & $29.6$ & $29.3$ & $25.6$ \\
        \hline
        \multicolumn{6}{c}{Viscous model $s=2$} \\
        SNe Ia & $2900$ & $0.335$ & $29.1$ & $56.6$ & $52.2$ \\
        OHD & $1650$ & $0.355$ & $26.9$ & $30.1$ & $32.0$ \\
        SNe Ia+OHD & $1650$ & $0.367$ & $23.6$ & $24.9$ & $28.3$ \\
    \end{tabular}
    \end{ruledtabular}
\end{table*}

\subsection{\label{sec:Results}Results and discussion}
\begin{figure*}
    \centering
    \includegraphics[scale = 0.36]{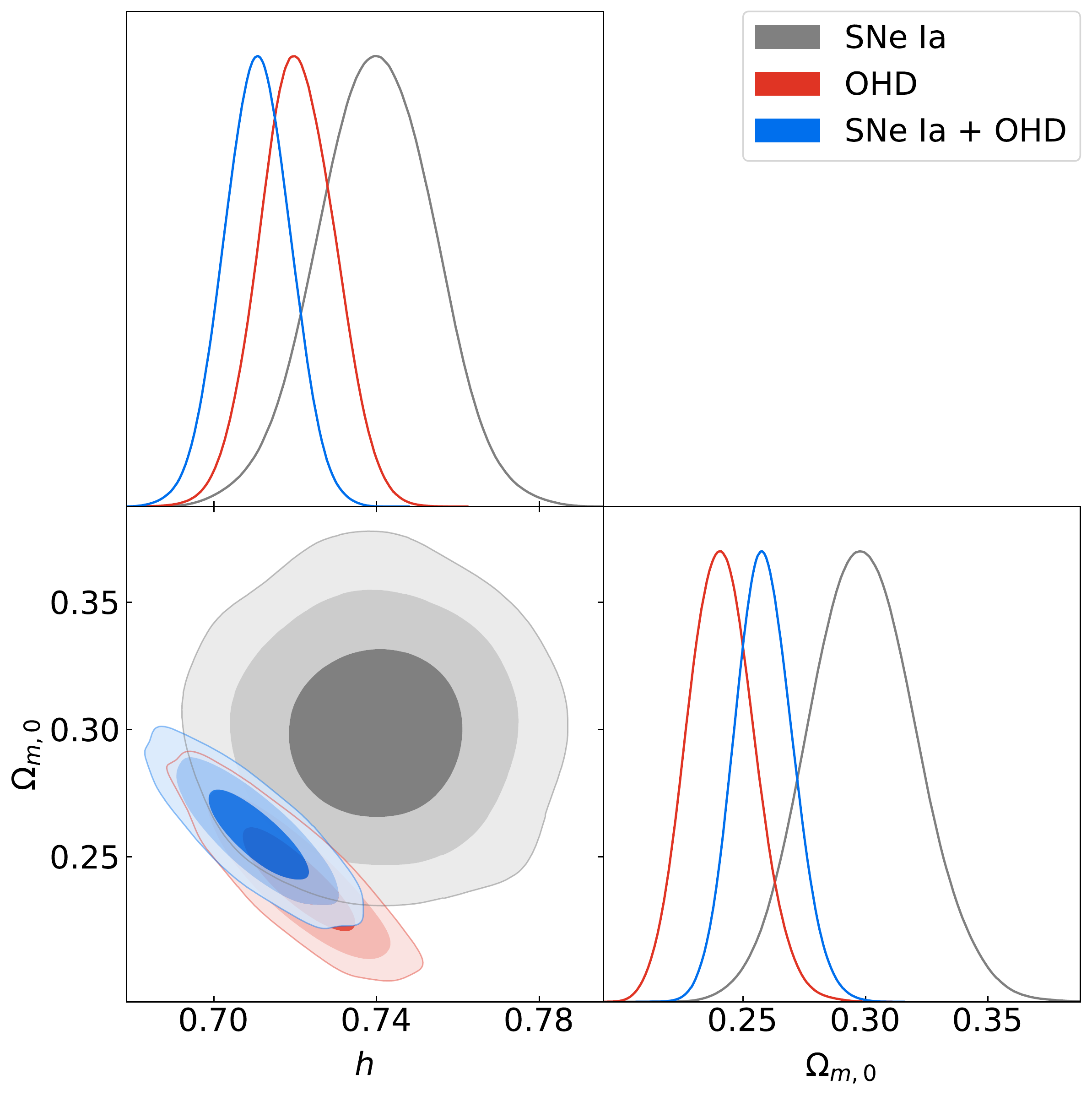}
    \caption{\label{fig:TriangleLambdaCDM} Joint and marginalized regions of the free parameters $h$ and $\Omega_{m,0}$ for the $\Lambda$CDM model, obtained in the MCMC analysis described in the Section \ref{Constraints} for the SNe Ia data, OHD, and in their joint analysis. The admissible joint regions correspond to $1\sigma(68.3\%)$, $2\sigma(95.5\%)$, and $3\sigma(99.7\%)$ of confidence level (CL), respectively. The best-fit value for each free parameter are presented in the Table \ref{tab:BestFit}.}
\end{figure*}

\begin{figure*}
    \centering
    \subfigure[\label{fig:TriangleViscouss-15} Joint and marginalized regions for the Viscous model with $s=-3/2$.]{\includegraphics[scale = 0.36]{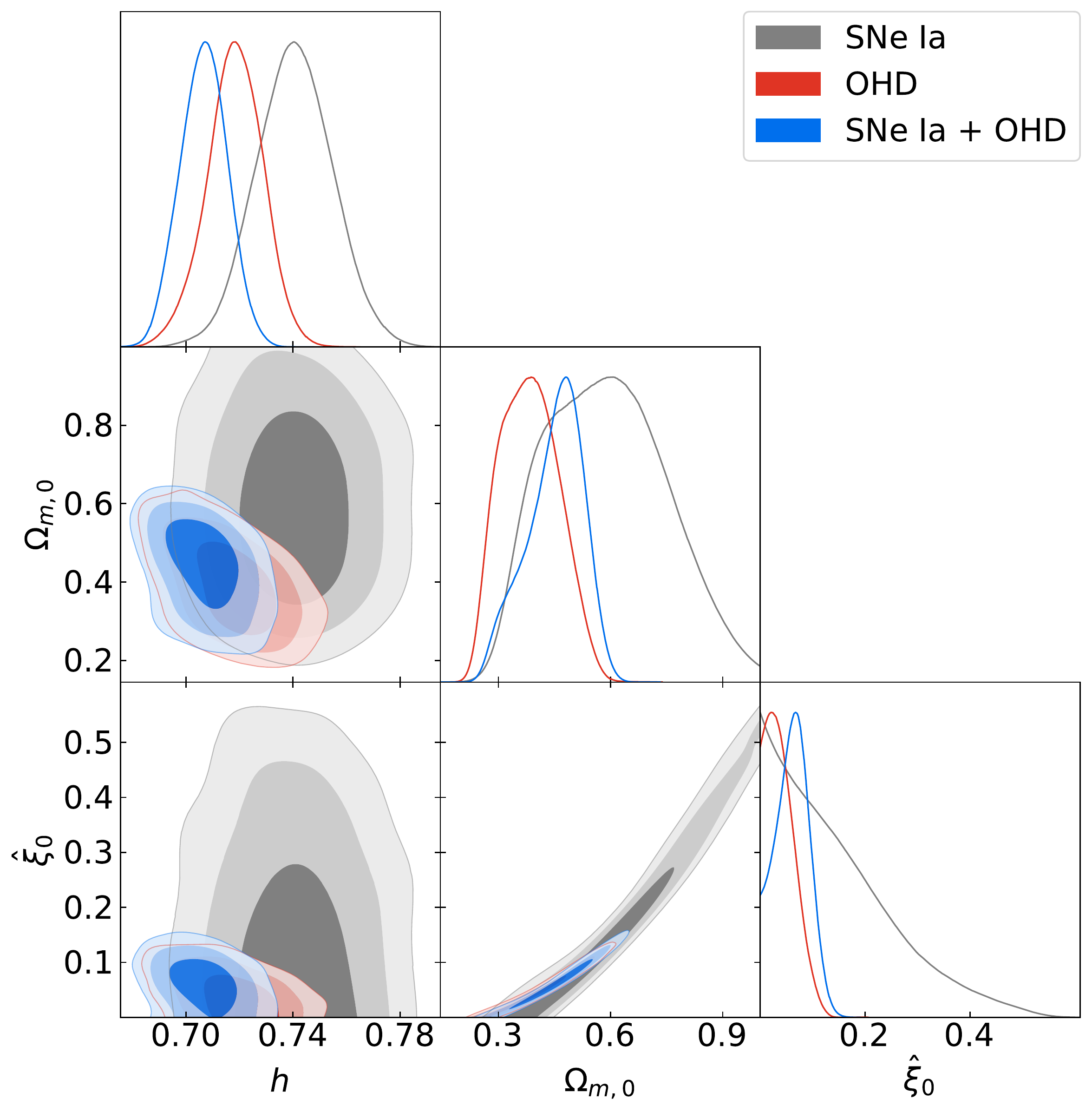}}
    \subfigure[\label{fig:TriangleViscouss-05} Joint and marginalized regions for the Viscous model with $s=-1/2$.]{\includegraphics[scale = 0.36]{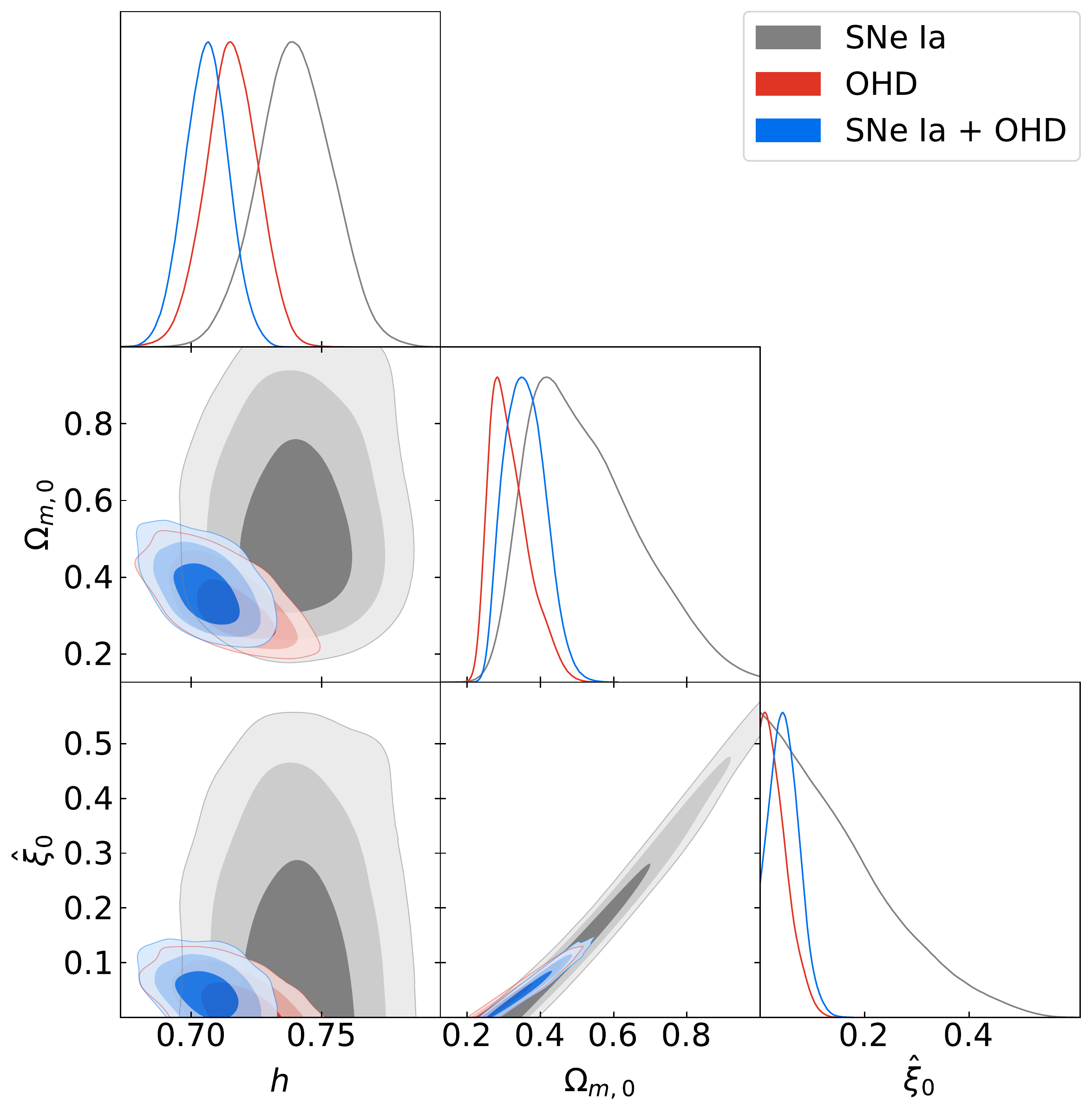}}
    \caption{\label{fig:TriangleViscousNegative} Joint and marginalized regions of the free parameters $h$, $\Omega_{m,0}$, and $\hat{\xi_{0}}$ for the Viscous model with $s=-3/2$ and $-1/2$, obtained in the MCMC analysis described in the Section \ref{Constraints} for the SNe Ia data, OHD, and in their joint analysis. The admissible joint regions correspond to $1\sigma(68.3\%)$, $2\sigma(95.5\%)$, and $3\sigma(99.7\%)$ of confidence level (CL), respectively. The best-fit value for each free parameter, for each case, are presented in the Table \ref{tab:BestFit}.}
\end{figure*}

\begin{figure*}
    \centering
    \subfigure[\label{fig:TriangleViscouss05} Joint and marginalized regions for the viscous model with $s=1/2$.]{\includegraphics[scale = 0.36]{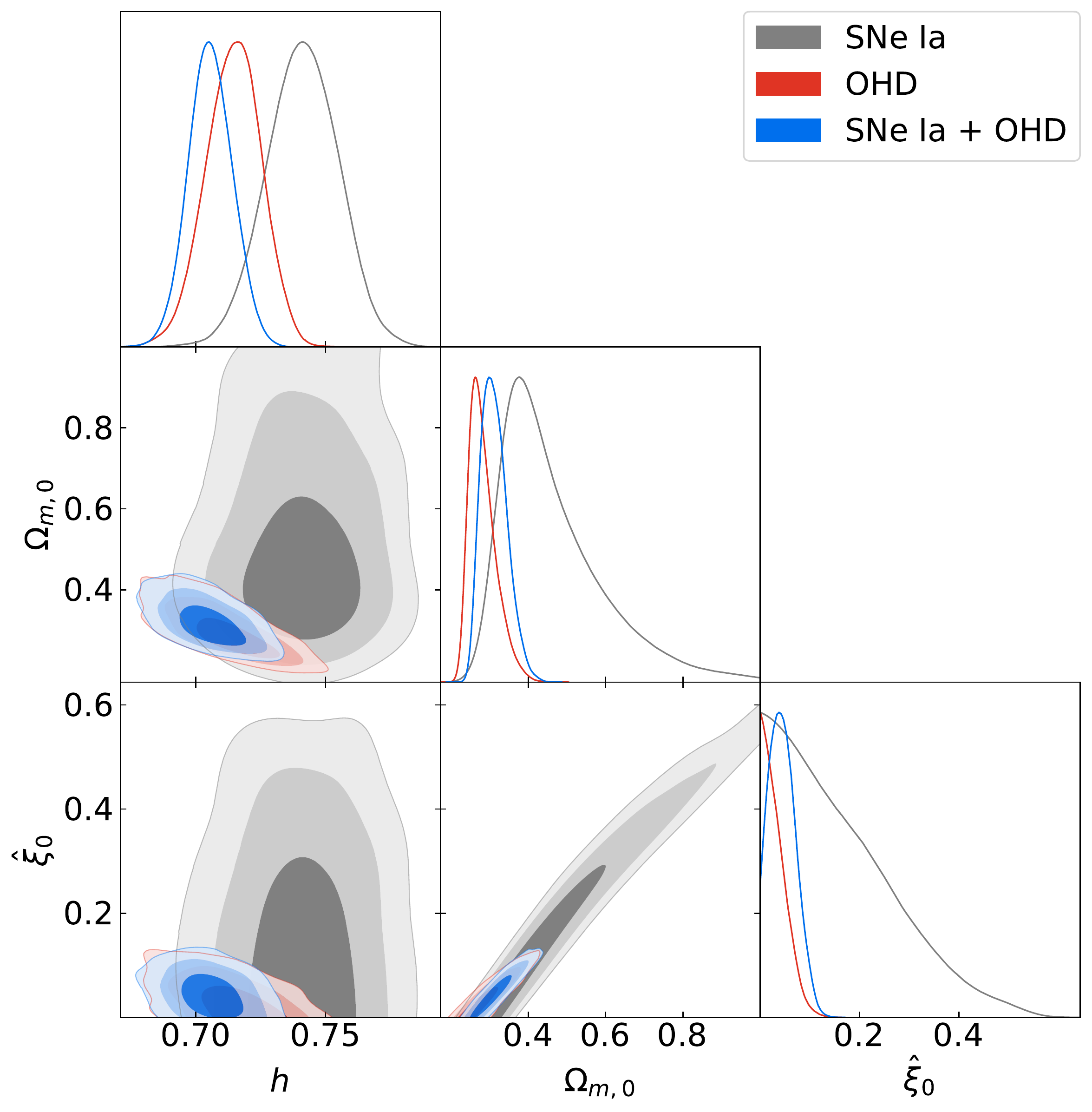}}
    \subfigure[\label{fig:TriangleViscouss1} Joint and marginalized regions for the viscous model with $s=1$.]{\includegraphics[scale = 0.36]{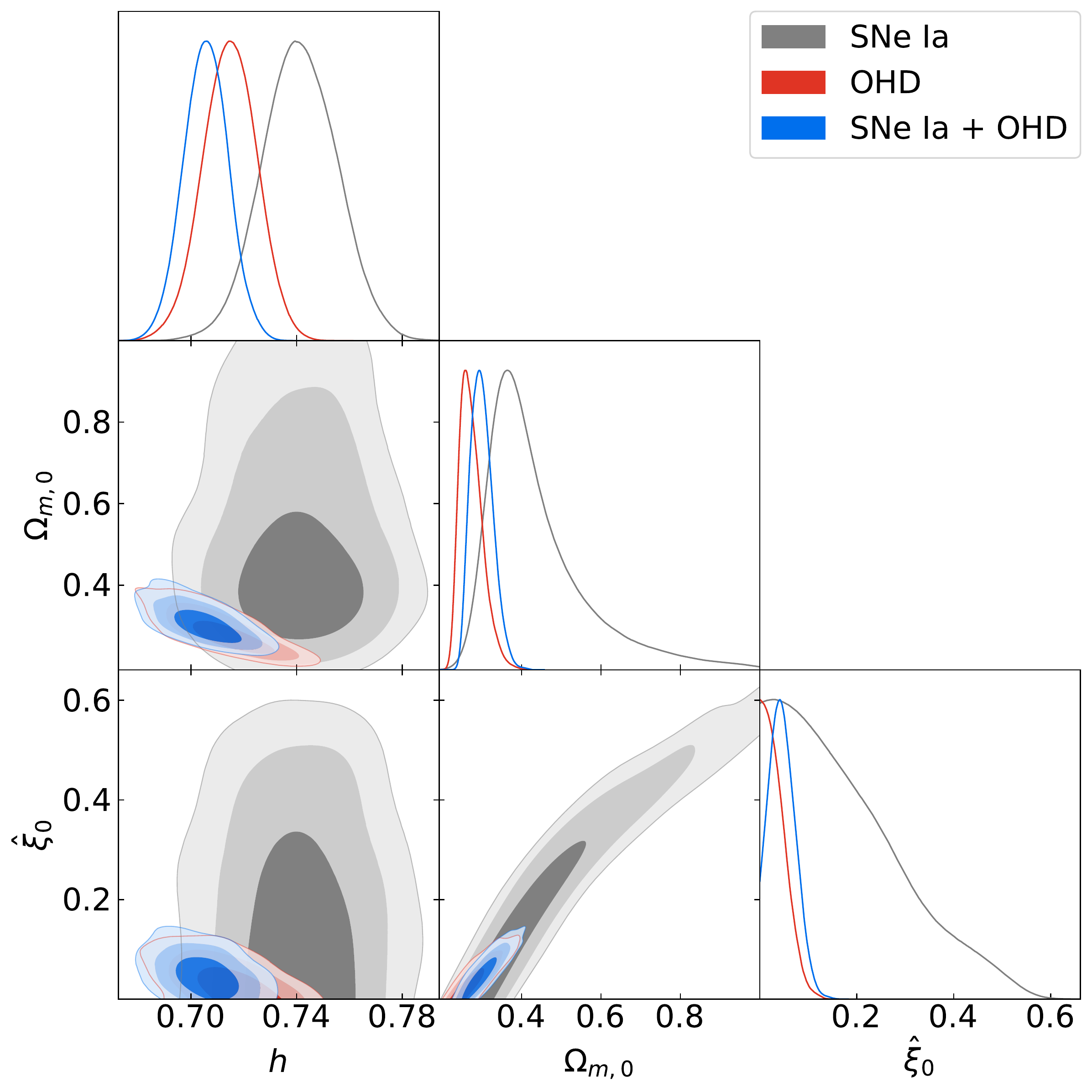}}
    \subfigure[\label{fig:TriangleViscouss15} Joint and marginalized regions for the viscous model with $s=3/2$.]{\includegraphics[scale = 0.36]{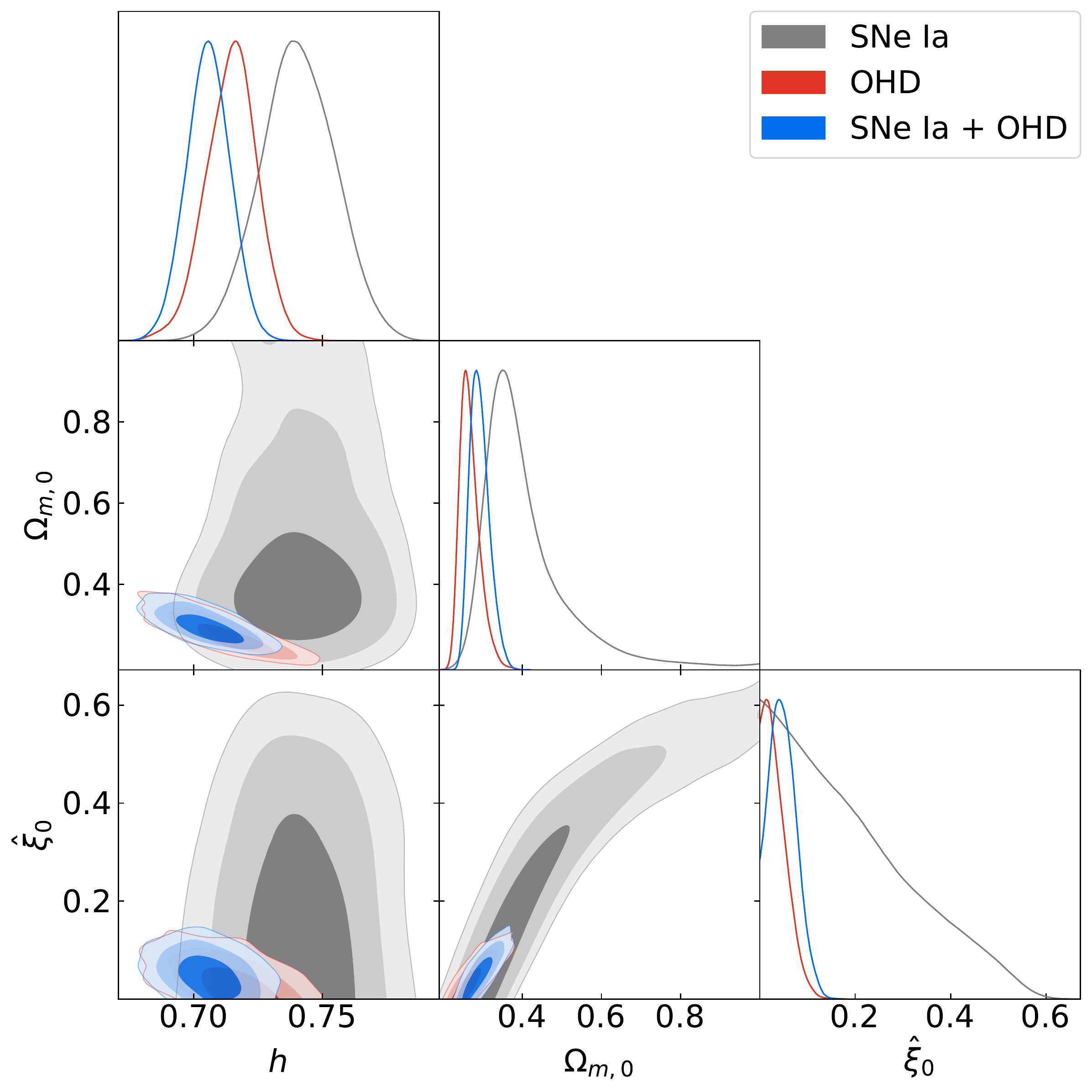}}
    \subfigure[\label{fig:TriangleViscouss2} Joint and marginalized regions for the viscous model with $s=2$.]{\includegraphics[scale = 0.36]{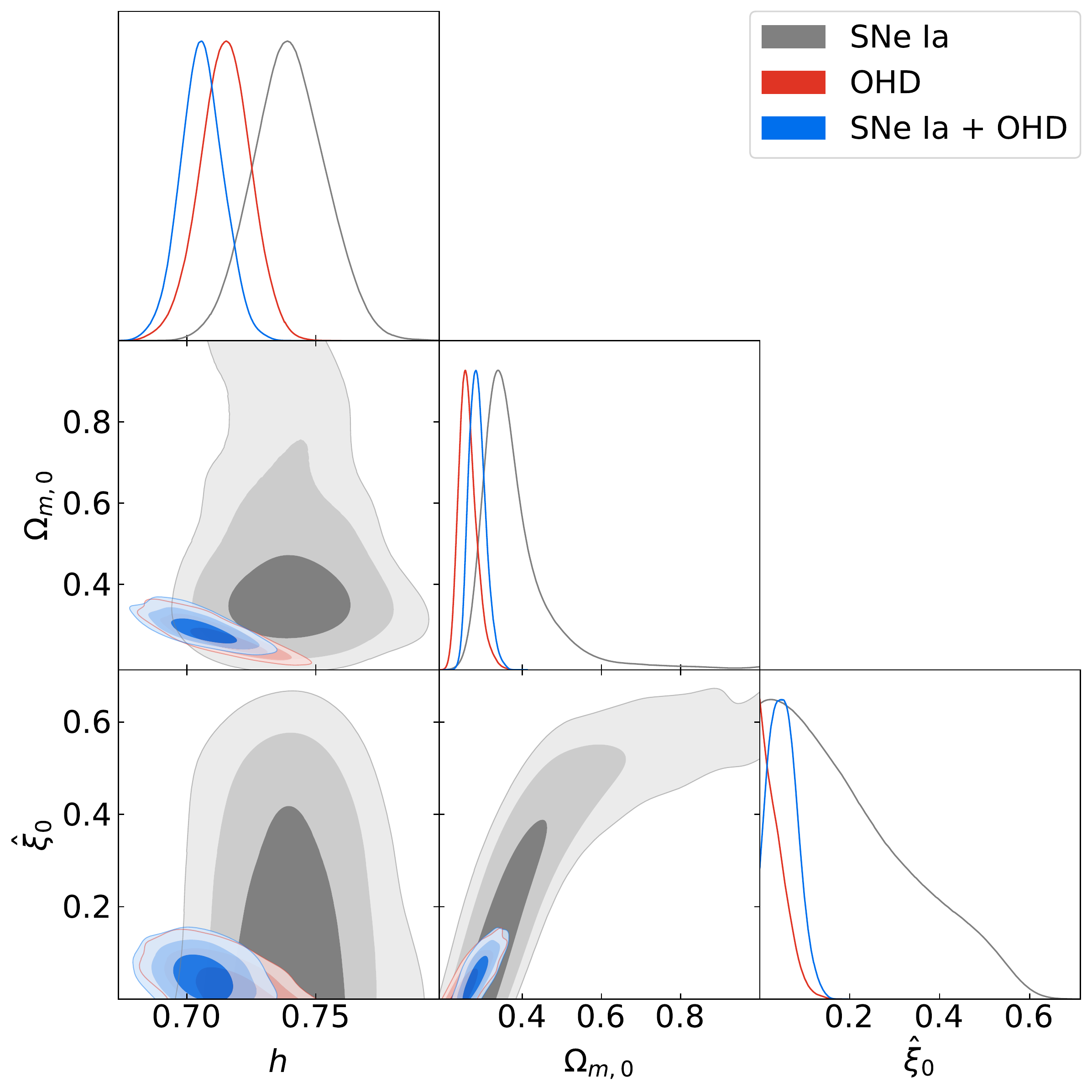}}
    \caption{\label{fig:TriangleViscousPositive} Joint and marginalized regions of the free parameters $h$, $\Omega_{m,0}$, and $\hat{\xi_{0}}$ for the viscous model with $s=1/2$, $1$, $3/2$, and $2$, obtained in the MCMC analysis described in the Section \ref{Constraints} for the SNe Ia data, OHD, and in their joint analysis. The admissible joint regions correspond to $1\sigma(68.3\%)$, $2\sigma(95.5\%)$, and $3\sigma(99.7\%)$ of confidence level (CL), respectively,
    with the viscous model displaying an even better fit. The best-fit value for each free parameter, for each case, are presented in the Table \ref{tab:BestFit}.}
\end{figure*}

The best-fit values for the free parameters $h$, $\Omega_{m,0}$, and $\hat{\xi}_{0}$ for the viscous model with $s=-3/2$, $-1/2$, $1/2$, $1$, $3/2$, and $2$, as well as the best-fit values for the free parameters $h$ and $\Omega_{m,0}$ for the $\Lambda$CDM model, are presented in Table \ref{tab:BestFit}. We also present their corresponding goodness-of-fit criteria, given by the values of $\chi_{\rm min}^{2}$ and BIC. The uncertainties shown correspond to $1\sigma(68.3\%)$ of confidence level (CL). In Figure \eqref{fig:TriangleLambdaCDM}, we depict the joint and marginalized regions of the free parameters $h$ and $\Omega_{m,0}$ for the $\Lambda$CDM model; while in Figures \ref{fig:TriangleViscouss-15}, \ref{fig:TriangleViscouss-05}, \ref{fig:TriangleViscouss05}, \ref{fig:TriangleViscouss1}, \ref{fig:TriangleViscouss15}, and \ref{fig:TriangleViscouss2} we do the same for $h$, $\Omega_{m,0}$, and $\hat{\xi}_{0}$ for the viscous model with $s=-3/2$, $-1/2$, $1/2$, $1$, $3/2$, and $2$, respectively. All the derived constraints were obtained by the MCMC analysis described in Section \ref{Constraints} for the SNe Ia data, OHD, and the joint analysis, as appropriate. The admissible joint regions correspond to $1\sigma$, $2\sigma(95.5\%)$, and $3\sigma(99.7\%)$ CL.

\begin{table*}
    \centering
    \caption{\label{tab:BestFit} Best-fit values for the free parameters $h$, $\Omega_{m,0}$, and $\hat{\xi}_{0}$ for the viscous model with $s=-3/2$, $-1/2$, $1/2$, $1$, $3/2$, and $2$; as well as their corresponding goodness-of-fit criteria, obtained in the MCMC analysis described in Section \ref{Constraints} for the SNe Ia data, OHD, and in their joint analysis. We also present the best-fit values for the free parameters $h$ and $\Omega_{m,0}$ for the $\Lambda$CDM model and their corresponding goodness-of-fit criteria as a further comparison. The uncertainties presented correspond to $1\sigma(68.3\%)$ of confidence level (CL).}
    \begin{ruledtabular}
    \begin{tabular}{cccccc}
        \multirow{2}{*}{Data} & \multicolumn{3}{c}{Best-fit values} & \multicolumn{2}{c}{Goodness-of-fit} \\
    \cline{2-6}
         & $h$ & $\Omega_{m,0}$ & $\hat{\xi}_{0}$ & $\chi_{\rm min}^{2}$ & BIC \\
    \hline
        \multicolumn{6}{c}{$\Lambda$CDM model} \\
        SNe Ia & $0.740_{-0.014}^{+0.014}$ & $0.299_{-0.022}^{+0.022}$ & $\cdots$ & $1026.9$ & $1040.8$ \\
        OHD & $0.720_{-0.009}^{+0.010}$ & $0.241_{-0.013}^{+0.014}$ & $\cdots$ & $28.6$ & $36.5$ \\
        SNe Ia+OHD & $0.711_{-0.008}^{+0.008}$ & $0.258_{-0.011}^{+0.012}$ & $\cdots$ & $1058.4$ & $1072.4$ \\
    \hline
    \multicolumn{6}{c}{Viscous model $s=-3/2$} \\
        SNe Ia & $0.740_{-0.015}^{+0.015}$ & $0.583_{-0.176}^{+0.170}$ & $0.122_{-0.089}^{+0.140}$ & $1027.4$ & $1048.2$ \\
        OHD & $0.718_{-0.010}^{+0.010}$ & $0.382_{-0.082}^{+0.085}$ & $0.037_{-0.025}^{+0.033}$ & $27.5$ & $39.3$ \\
        SNe Ia+OHD & $0.707_{-0.009}^{+0.009}$ & $0.460_{-0.095}^{+0.064}$ & $0.062_{-0.035}^{+0.031}$ & $1055.0$ & $1076.0$ \\
    \hline
        \multicolumn{6}{c}{Viscous model $s=-1/2$} \\
        SNe Ia & $0.740_{-0.014}^{+0.015}$ & $0.504_{-0.139}^{+0.184}$ & $0.126_{-0.090}^{+0.143}$ & $1027.6$ & $1048.5$ \\
        OHD & $0.715_{-0.010}^{+0.010}$ & $0.306_{-0.045}^{+0.064}$ & $0.029_{-0.020}^{+0.031}$ & $28.0$ & $39.8$ \\
        SNe Ia+OHD & $0.706_{-0.009}^{+0.008}$ & $0.355_{-0.054}^{+0.058}$ & $0.046_{-0.026}^{+0.029}$ & $1055.2$ & $1076.2$ \\
    \hline
    \multicolumn{6}{c}{Viscous model $s=1/2$} \\
        SNe Ia & $0.741_{-0.014}^{+0.014}$ & $0.424_{-0.093}^{+0.168}$ & $0.136_{-0.096}^{+0.149}$ & $1027.8$ & $1048.7$ \\
        OHD & $0.715_{-0.011}^{+0.010}$ & $0.275_{-0.028}^{+0.042}$ & $0.027_{-0.019}^{+0.029}$ & $28.1$ & $39.9$ \\
        SNe Ia+OHD & $0.706_{-0.008}^{+0.009}$ & $0.309_{-0.032}^{+0.037}$ & $0.044_{-0.026}^{+0.029}$ & $1055.2$ & $1076.2$ \\
    \hline
    \multicolumn{6}{c}{Viscous model $s=1$} \\
        SNe Ia & $0.741_{-0.014}^{+0.014}$ & $0.395_{-0.073}^{+0.154}$ & $0.151_{-0.105}^{+0.157}$ & $1027.9$ & $1048.8$ \\
        OHD & $0.715_{-0.010}^{+0.010}$ & $0.269_{-0.025}^{+0.033}$ & $0.030_{-0.021}^{+0.029}$ & $28.1$ & $39.9$ \\
        SNe Ia+OHD & $0.706_{-0.009}^{+0.008}$ & $0.298_{-0.027}^{+0.031}$ & $0.045_{-0.025}^{+0.030}$ & $1055.2$ & $1076.2$ \\
    \hline
    \multicolumn{6}{c}{Viscous model $s=3/2$} \\
        SNe Ia & $0.740_{-0.014}^{+0.015}$ & $0.368_{-0.057}^{+0.131}$ & $0.161_{-0.117}^{+0.183}$ & $1027.9$ & $1048.8$ \\
        OHD & $0.715_{-0.011}^{+0.009}$ & $0.264_{-0.020}^{+0.029}$ & $0.030_{-0.020}^{+0.030}$ & $28.1$ & $39.9$ \\
        SNe Ia+OHD & $0.706_{-0.009}^{+0.008}$ & $0.290_{-0.022}^{+0.027}$ & $0.047_{-0.027}^{+0.031}$ & $1055.2$ & $1076.2$ \\
    \hline
    \multicolumn{6}{c}{Viscous model $s=2$} \\
        SNe Ia & $0.740_{-0.014}^{+0.015}$ & $0.350_{-0.044}^{+0.098}$ & $0.174_{-0.124}^{+0.195}$ & $1027.9$ & $1048.7$ \\
        OHD & $0.715_{-0.010}^{+0.010}$ & $0.260_{-0.020}^{+0.026}$ & $0.030_{-0.022}^{+0.033}$ & $28.0$ & $39.8$ \\
        SNe Ia+OHD & $0.706_{-0.008}^{+0.009}$ & $0.285_{-0.020}^{+0.022}$ & $0.051_{-0.030}^{+0.033}$ & $1055.1$ & $1076.1$ \\
    \end{tabular}
    \end{ruledtabular}
\end{table*}

Focusing particularly on the values of $\chi_{\rm min}^{2}$ obtained in Table \ref{tab:BestFit}, we can see that for the SNe Ia data the $\Lambda$CDM model has a slightly better fit than the viscous model independently of the value of $s$; while for the OHD the viscous model presents a slightly better fit than the $\Lambda$CDM model, again independently of the value of $s$, though with a more remarkable difference for the case $s=-3/2$. However, in their joint analysis the differences between both models, for each selected case of $s$, are more distinctive, with the viscous model displaying an even better fit.

In spite of the above encouraging feature,  there is  statistically strong evidence against the viscous model compared to the $\Lambda$CDM model, based only on the SNe Ia data. While using  OHD, and their joint analysis, there is evidence against the viscous model. Interestingly, the statistical significance is quite independent of the value of $s$. For instance, we find differences of BIC within the range $3.6$ and $3.8$ for the best and worse cases in the joint analysis, respectively. We argue then that such differences in BIC between both models cannot be considered as a lapidary proof against the viscous model. Therefore, the inclusion of more data is a compulsory strategy to assess which model is statistically preferred. This is a challenging program which will be addressed in a future work.

When comparing the values of $\chi_{\rm min}^{2}$ and BIC for the different elections of $s$, very slightly differences are appreciable among  them, which clearly means that there is no preference for one specific value of $s$ from the background data used in our MCMC analysis. This is an expected result,  which agrees  with the  corresponding one found in the dynamical system analysis described in the previous section. 

It is important to mention that, a careful examination of the best-fit values for the free parameters of the viscous model leads to a more compelling conclusion about the election of $s$. Before explaining this in detail, first notice that another element to be considered in the analysis is the fact that the viscous model has an extra free parameter that make definitely a difference in the estimations. Following this argument, we can expect in principle to find best-fit values for the free parameter $\Omega_{m,0}$ for the viscous model that are closer to the ones of the reference model. According to this discussion and considering the joint analysis only, we can see that the best-fit values for $\Omega_{m,0}$ for the cases $s=-3/2$ and $-1/2$ are further away from the one obtained for the $\Lambda$CDM. More importantly, the best-fit value of $\Omega_{m,0}$ for the viscous model approaches to the corresponding one of the $\Lambda$CDM model for positive and larger $s$. This result is a consequence of the functional form used for the bulk viscosity given by Eqn.\eqref{sec2:eqn3}, expression that leads to $\xi\to 0$ when  $s\to\infty$ for $\Omega_{m,0}\neq 1$ (assuming  of course that $H$ does not grow too fast in the redshift range considered), thus the viscous model reduces naturally to the $\Lambda$CDM model. Therefore, for the background data used in our MCMC analysis, namely, SNe Ia data and OHD, we have evidence on positive values of $s$ over negative ones.

\begin{figure*}
    \centering
    \includegraphics[scale = 0.43]{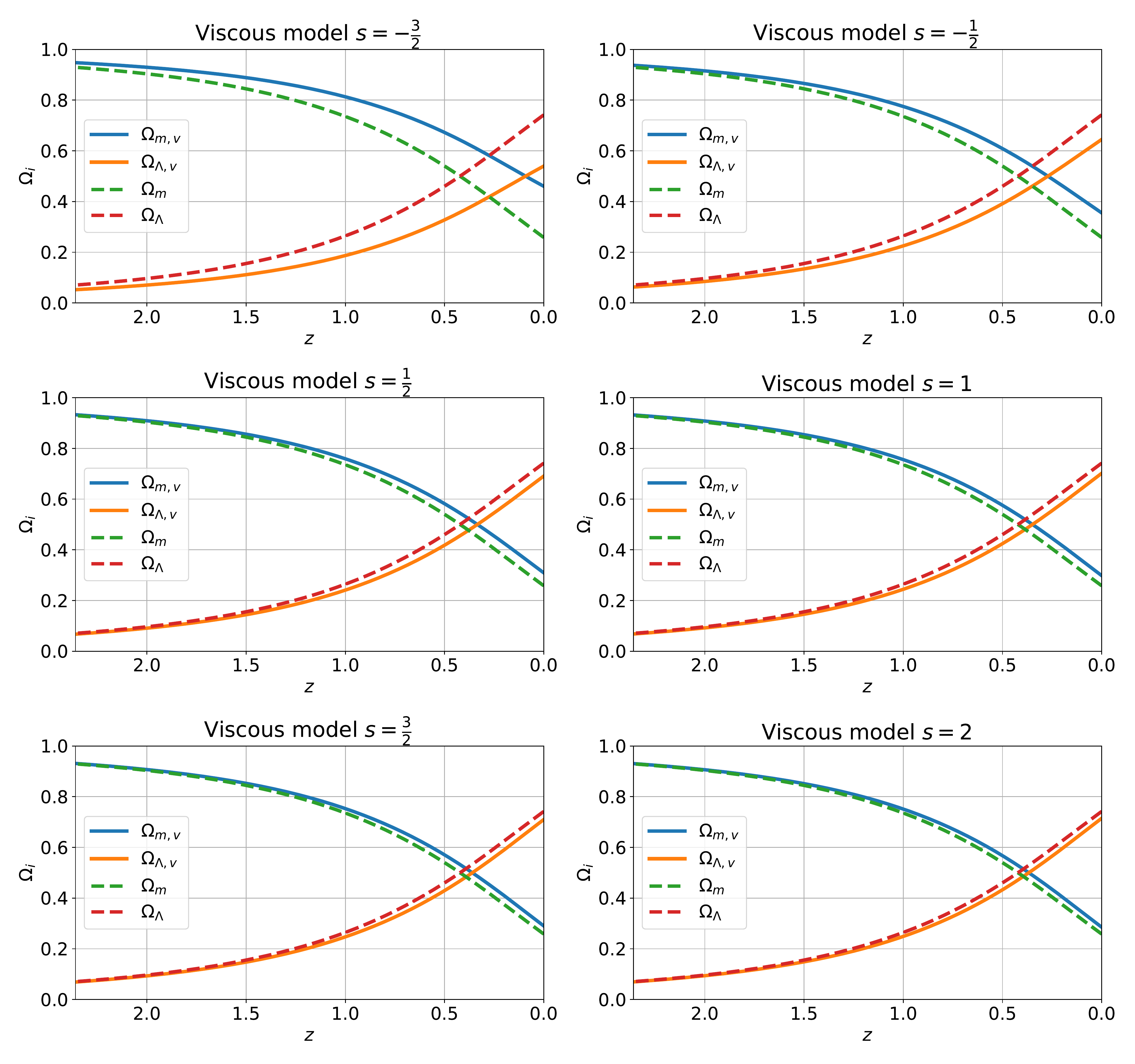} \
    \caption{\label{fig:Density} Evolution of the density parameters $\Omega_{i}$ against the redshift $z$, using the best-fit values for the joint analysis presented in Table \ref{tab:BestFit} as initial conditions at $z=0$. The solid lines correspond to the density parameters $\Omega_{i,v}$ for the viscous model, obtained from the numerical integration of Eqn. \eqref{ConstrainModel}; while the dashed lines correspond to the density parameters $\Omega_{i}$ for the $\Lambda$CDM model, obtained from Eqn. \eqref{LCDM}. Here $i$ stands for $m$ (DM) and $\Lambda$ (DE). The x-axis is presented in the range $0\leq z\leq 2.36$, being $z=0$ the current time and $z=2.36$ the highest redshift in the combined OHD and Pantheon samples.}
\end{figure*}

\begin{figure*}
    \centering
    \includegraphics[scale = 0.43]{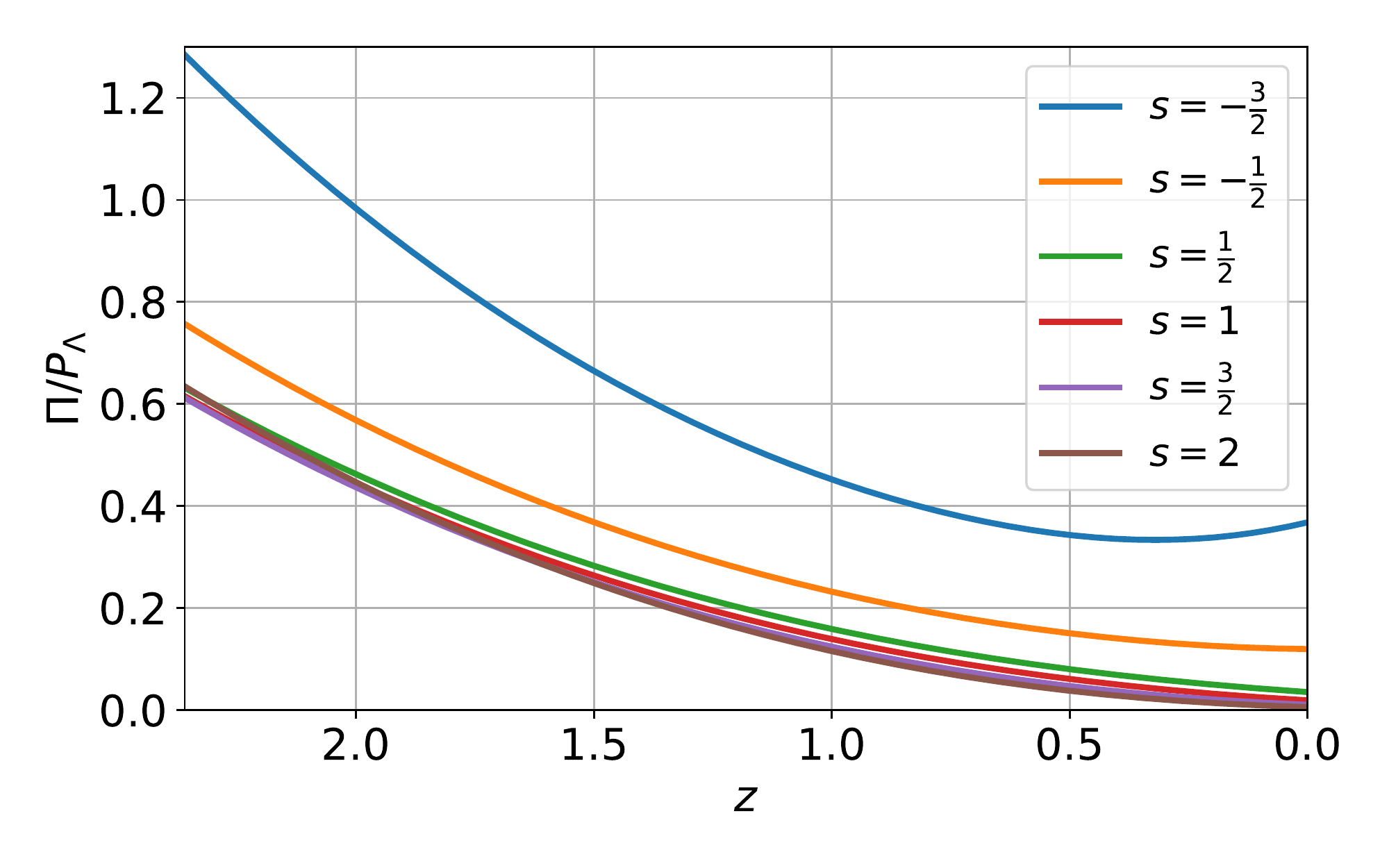}
    \includegraphics[scale = 0.43]{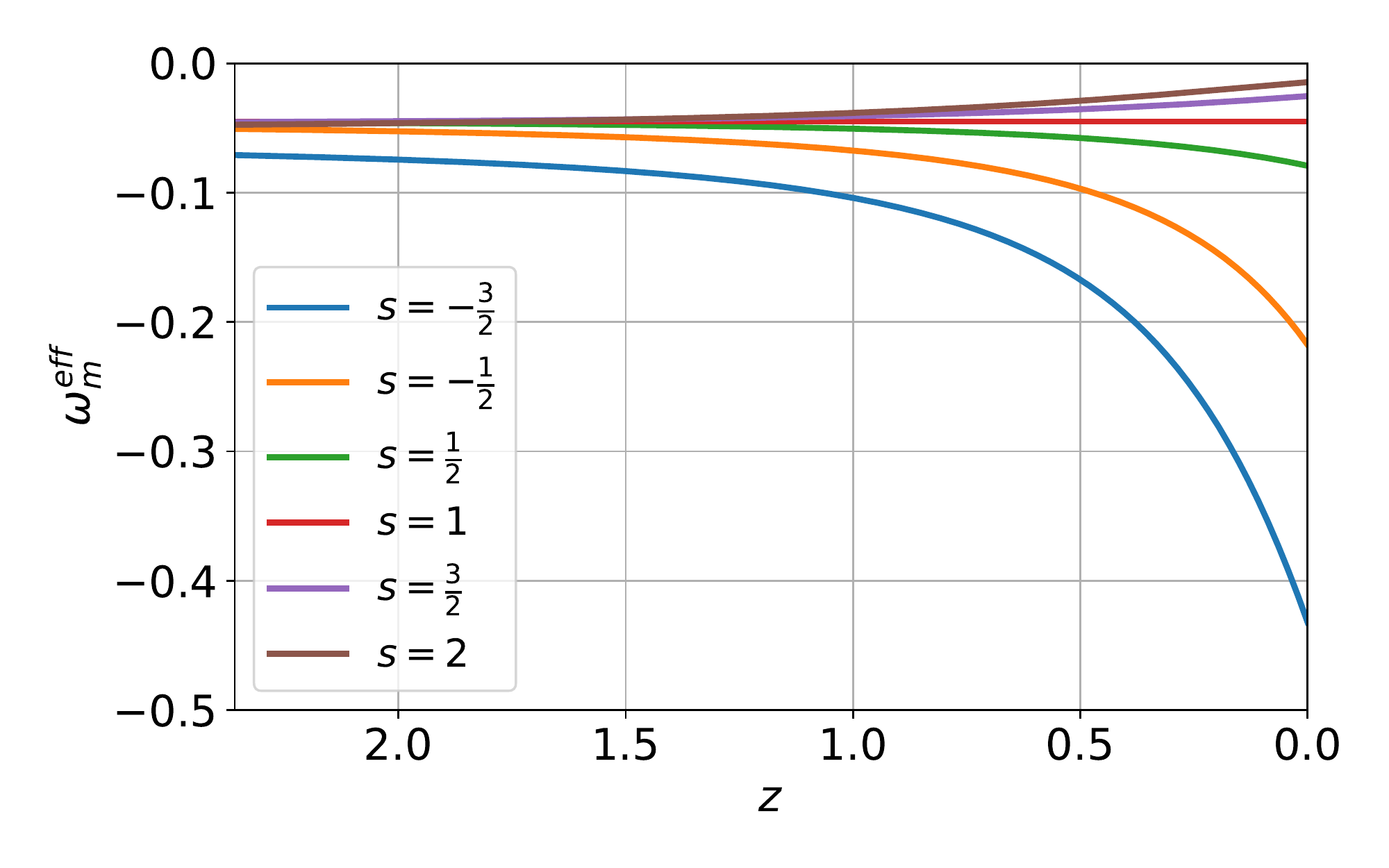}
    \caption{\label{fig:Pressure-omegaeffe} \textit{Left panel}: Evolution of the ratio $\Pi/P_{\Lambda}$ against the redshift $z$ obtained from Eqn. \eqref{Bulkpressure}. \textit{Right panel}: Evolution of the effective EoS for DM, $\omega_{m}^{\rm eff}$, against the redshift $z$ from Eqn. \eqref{eqn:EoSDM}. In both cases, we consider the best-fit values for the joint analysis presented in Table \ref{tab:BestFit} as initial conditions at $z=0$. The x-axis is presented in the range $0\leq z\leq 2.36$, being $z=0$ the current time and $z=2.36$ the highest redshift in the combined OHD and Pantheon samples.}
\end{figure*}

Going deeper into the above, in Figure \ref{fig:Density} we depict the evolution of the density parameters $\Omega_{i}$, where i stands for $m$ (DM) and $\Lambda$ (DE), against the redshift $z$, for the viscous model with $s=-3/2$, $-1/2$, $1/2$, $1$, $3/2$, and $2$, and for the $\Lambda$CDM model. We use as initial conditions the best-fit values for the joint analysis presented in Table \ref{tab:BestFit}. According to this figure, the evolution of the density parameters of DM and DE for the viscous model looks more similarly than their $\Lambda$CDM counterparts for large positive values of $s$. On the contrary, for large negative s-values, the viscous model behaves  differently from the $\Lambda$CDM model, in particular during the transition towards present times. Again, this is a direct consequence of the functional form considered for the bulk viscosity, which has the remarkable feature that, in the limit $\hat{\xi}_{0}\to 0$ or $s\to\infty$, the $\Lambda$CDM model is recovered. This is clearly a natural limit of the viscous model. Following the same reasoning, negative values of $s$ lead to larger bulk viscosity coefficients, as can be derived from Eqn. \eqref{sec2:eqn3}. As a possible implication, the bulk viscosity pressure may exert, depending on the $s$ value, an extra negative pressure that can accelerate the universe expansion\footnote{Another physical interpretation can be given in terms of an effective pressure for DE that arises due to the contribution of the bulk viscosity pressure in the acceleration equation Eqn.~(\ref{sec2:eqn1}). Accordingly, deviation of $\omega_{\rm DE}=-1$ is possible, leading to an artificial phantom scenario.}. To see this better, we depict the evolution of the ratio $\Pi/P_{\Lambda}$ as a function of the redshift $z$ in the left panel of Figure \ref{fig:Pressure-omegaeffe}, according to the expression
%
\begin{equation}\label{Bulkpressure}
    \frac{\Pi}{p_{\Lambda}}=\frac{\hat{\xi}_{0}}{(1-\Omega_{m,0})}E^{2}\Omega_{m}^{s},
\end{equation}
and in the right panel of Figure \ref{fig:Pressure-omegaeffe}, the evolution of the effective EoS for DM, $\omega_{m}^{\rm eff}=P_{m}^{\rm eff}/\rho_{m}$, according to the expression 
\begin{equation}
    \omega_{m}^{\rm eff}=\frac{\Pi}{\rho_{m}}=-\hat{\xi}_{0}\Omega_{m}^{s-1}\label{eqn:EoSDM},
\end{equation}
for the viscous model with $s=-3/2$, $-1/2$, $1/2$, $1$, $3/2$, and $2$, using the same initial conditions as before. In Eqn.~(\ref{Bulkpressure}) the dimensionless Hubble parameter $E\equiv H/H_{0}$ has been introduced. Notice that for $s=-1/2$ and $s=-3/2$ the contribution of the bulk viscous pressure is not negligible at the present time ($z=0$), contrary to their $s$-positive counterparts. Even more, the dissipative pressure for the case $s=-3/2$ shows larger differences than the other cases, exceeding shortly after $z=2$ the DE pressure (see left panel of Figure \ref{fig:Pressure-omegaeffe}). Hence, this plot gives important clues on when the dissipative pressure dominates over the DE pressure. Nevertheless, this does not show in reality at which point in the background evolution the dissipative pressure contributes more significantly.
To get then a more realistic and transparent interpretation of the bulk viscosity effect it is necessary to observe, instead, the behavior of $\omega_{m}^{\rm eff}$ (see right panel of Figure \ref{fig:Pressure-omegaeffe}). This tells us, after careful examination, that the more important contribution occurs at very low redshifts, where $\omega_{m}^{\rm eff}$ can take values even lower than $-1/3$. Interestingly, this extra acceleration due to the bulk viscous pressure is compensated in the fit with large best-fit values for $\Omega_{m,0}$. It implies of course to have smaller energy density parameters associated to the cosmological constant\footnote{In the extreme case of large bulk viscosity values, which is achieved for large negative values of $s$, the presence of dark energy is not required to accelerate the expansion. This physical situation corresponds indeed to an unified dark fluid scenario, and it is another natural convergence of this viscous model.}, as can be  inferred in Table \ref{tab:BestFit}.

As a general feature we obtain $\omega_{m}^{\rm eff}<0$ for any value of $s$. This is a noticeable prediction of the viscous model that deserves further attention either as an extra source of generating the accelerated expansion, or as an alternative model to describe the DM component beyond the standard cold DM.

Other anomalous effect due to the large contribution of the bulk viscosity is the abrupt displacement of the redshift of matter-DE equality, $z_{\rm eq}$, towards the present time, in comparison to the reference model. The larger difference is presented for the case $s=-3/2$, with a shift numerically obtained of $\Delta z_{\rm eq}\approx0.33$  (see top panels of Figure \ref{fig:Density}). As $s$ increases, such differences are clearly attenuated (see bottom panels). The reason of this mismatch is precisely due to the distinctive evolution of the
energy density parameters, in particular  during the transition period, that is, after full matter dominated period. This affects the onset of the accelerated expansion, which is set in a model independent way, according to observations of the Hubble parameter provided by the Baryon Oscillation Spectroscopic Survey Data Release 9, around $z=0.64$ \cite{Moresco:2016mzx}. Therefore, we have preliminary evidence against negative values of $s$ in the novel parametrization for the bulk viscosity Eqn. \eqref{sec2:eqn3}, and preference for positive values based on the best-fits and on the well-behaved cosmological evolution.


\section{Conclusions}\label{conclusions}

We have investigated the cosmological implications of a new parameterization for the bulk viscosity which includes most of the well-known cases within the framework of the Eckart's theory. 
Our viscous model consists of a non-trivial combination of:  
i) the Hubble parameter,
and 
ii) the dark matter energy density, 
of the form $\xi\sim H^{1-2s} \rho_{m}^{s}$, for an arbitrary $s$-exponent. 
This functional form clearly leads to a direct coupling of dark matter to the other components of the universe whereby the bulk viscosity effects, as dictated by the dynamical system, are effective only when dark matter is (sub) dominant, provided $s>0$.  

Our proposal is quite general in the sense that encloses naturally the well-known cases of phenomenological parameterizations for the bulk viscosity $s= \{ 0, 1/2 \}$, and more importantly, it allows the existence of new viscous solutions for any arbitrary values of the $s$-exponent. 

The methodology we have implemented in this work is based mainly on three approaches: 
i) a systematic symbolic programming for an arbitrary $s$-exponent that resulted in Tables \ref{M1:critical points} and \ref{M1:eigenvalues}  of section \ref{sec:dynsys:symbolic}, 
ii) a detailed analytical dynamical system analysis performed in section \ref{sec:dynsys:analytical}, and 
iii) an best-fit data analysis and parameters estimations through the Monte Carlo method in section \ref{sec:constraints}. We have found full consistency between both independent approaches carried out in i) and ii), and whose constraints derived from the stability criteria were used as an input in the subsequent cosmological constraints. 

Concretely based on approaches i) and ii), we have found one-parameter family of de-Sitter-like asymptotic solutions with non-vanishing bulk viscosity that can be classified into two large groups: positive integers (excluding $s\neq1$) and half-integer values of $s$. These solutions are clearly distinguishable from each other due to the possible presence of the bulk viscosity coefficient during the radiation dominated period for all half-integer values of $s$. They share, however, similarities at late times where energy density parameters are of the appealing form  $\Omega_{\Lambda}=1-\hat{\xi}_{0}^{\frac{1}{1-s}}$ and $\Omega_{m}=\hat{\xi}_{0}^{\frac{1}{1-s}}$ for $s\neq1$. 

On the other hand, the presence of these de-Sitter-like asymptotic behaviors leads to interesting cosmological scenarios in the context of alleviate the so called “coincidence problem”. Indeed, since the cosmic  densities of both dark component are of the same order today, despite their different decreasing rates, it is an open question to understand why this behavior occurs. For example in the case $s=-3/2$, and despite the fact that for positive and large $s$, the corresponding best-fit value approaches closely to the one inferred for the $\Lambda$CDM model, we can see from Table I that in the case of the (Bc) point, the asymptotic future for both DE and DM densities tend to non-zero constants, being $1-\hat{\xi}_0^{2/5}$ and $\hat{\xi}_0^{2/5}$, respectively. Using the joint best-fit, given in this case by $\hat{\xi}_0 = 0.062$, we obtain that $\Omega_{\Lambda} = 2\Omega_m $, which corresponds to a universe with a de Sitter expansion in the far future but with a non-zero matter density. These behavior also occurs  for the cases $s=-1/2, 1,2$. In other words, we have found that viscosity acts as mechanism that can lead to similar energy densities of both dark fluids during the cosmic evolution.

Demanding complete cosmological dynamics, solutions corresponding to both negative integers 
values of $s$ and the zero case, i.e. $\xi\propto H$, are simply discarded because they can not provide radiation domination periods as can be read from Table \ref{M1:critical points}. This physical criterion applies also on unified dark fluid models with $\xi\propto H$. Despite that they have been extensively used as cosmological models, they must be clearly ruled out. This inference represents an important outcome of our parametrization using dynamical system analysis. This is also consistent with the result reported in Ref.~\cite{sasidharan2016phase}.
For other values of $s$, radiation domination is affected in a similar fashion than explained above through the viscous dissipation effect. Hence, these general solutions represent a one-viscosity parameter family of minimal extensions to the $\Lambda$CDM model.

In short, section \ref{sec:dynsys:symbolic} focused primarily on the cosmological implications of the viscous models and thorough discussion of their qualitative features. Section \ref{sec:dynsys:analytical}, on the contrary, focused on an analytical treatment of the fixed points for some $s$-exponents to explain formally the pattern structure of the fixed points found in the previous section. Successful and robust explanation of why a radiation point is absent for non-positive integer
values of $s$ was addressed with the aid of complex variable analysis. This treatment also allowed to explain the associated degeneracy of fixed points for rational $s$-values, which is associated to the multi-valued  exponential factors appearing in the expressions for the fixed points. This analytical treatment has the potential of finding all the existing stationary points for an arbitrary $s$-exponent. It is worthwhile mentioning that all our conclusions remain valid beyond the linear analysis. 

This work has been carry out in the framework of Eckart's theory but can be formulated in the causal IS formalism, following similar methods to the ones used in the present article. This is a challenging task to be performed in the future. Another phenomenological perspective of this work is the study of our novel parametrization in unified dark fluid scenarios in order to investigate whether current tensions rooted in the $\Lambda$CDM model can be overcome.

As a complementary and more robust test to assess the cosmological viability of this class of viscous models, cosmological constraints based on the MCMC method have been performed. Derived conclusions from this analysis are listed as follows: 
\begin{itemize}
    \item There is not preference for some specific viscous model when the background data is used in this analysis as it can be seen from the column Goodness-of-fit criteria of Table \ref{tab:BestFit}. This is in complete agreement with the dynamical system behavior which shows similarities and degeneracies between half-integers and positive integers values of the exponent $s$. The inclusion of cosmological data at high-redshift could in principle shed some light on this issue which appears when dealing with late-time data. 
    
    \item This first observational scrutiny tells us, based on the goodness-of-fit criteria for the joint analysis (SNe Ia data+OHD), that the $\chi_{\rm min}^{2}$ value obtained for each viscous model presents a better value than the corresponding one for the $\Lambda$CDM model. This conclusion holds for each selected case of $s$.
    
 \item Based on the joint analysis we have found statistical evidence against the viscous model in the whole range of considered $s$-values. Still, the BIC values are not a conclusive proof against the viscous model, whereby the inclusion of further data would be required to assess more certainly on which model is statistically preferred.

\item The analysis of the best-fit value of $\Omega_{m,0}$ gives a first assessment of which $s$ value is more suitable. Concretely, for positive and larger $s$,  the corresponding best-fit value approaches to the one inferred for the $\Lambda$CDM model. In other words, we have found evidence from data in favor of positive values of $s$ over negative ones.
    
\item The best-fit value of $\hat{\xi}_{0}$ is $\sim\mathcal{O}(10^{-2})$, which is larger compared with previous inferences derived from Large Scale Structure (LSS) observations $\hat{\xi}_{0}\sim\mathcal{O}(10^{-6})$  \cite{remedyforplankanlssdata}. It is worthwhile emphasizing however that our estimations for the bulk viscosity coefficient can not be compared with those values from the literature for two major reasons: first, the parametrization used in our manuscript is different from the ones studied in other works ($\xi=\rm const.$, $\xi\propto H$ and $\xi\propto \rho_{m}^{1/2}$) and aims to describe  the  bulk viscosity in  the DM  universe component (not for instance in DE), hence  a  direct comparison is therefore misleading. Second, the data set used in our work comes exclusively from the late-time phase of the universe and not from LSS. However, could be possible that once LSS data is considered in further analysis the bulk viscosity coefficient should be as small as the one inferred, for instance, in Ref.~\cite{remedyforplankanlssdata} where $\xi=\rm const.$ is assumed. Hence, our current estimations can be considered as marginal values and could be improved with the help of LSS data.

\item One  may  in  principle  compare  the  viscous  model with $s=0$ ($\xi\propto H$) with current inferences, however, as we stated based on dynamical system, this model must be ruled out because it can not provide a radiation domination period. This is the reason why we have not included it in our data analysis.
    
\end{itemize}
    
Some final inferences from the cosmological background evolution within the redshift range explored are as follows. When $s$ becomes increases in the negative axis, the impact of bulk viscosity on the energy density parameters becomes more important. On the contrary, when $s$ achieves increasing positive values, the behavior of $\Omega_i$ approaches to the one of the $\Lambda$CDM model (see Fig. \eqref{fig:Density}). Thus, we get viscous scenarios that effectively look like the $\Lambda$CDM solution at late times, nevertheless they could be different in other cosmological epochs. This appealing feature offers a rich phenomenological perspective to be investigated for instance in the context of structure formation and in the early Universe.

Moreover, dissipative pressure decays at late times while gives rise to large values at redshifts when DM is dominant. This is a physical desired property due to the dependence on the DM energy density.
In particular, the case $s=-3/2$ shows significant differences with respect to the $\Lambda$CDM model that can be presumably problematic due to the abrupt displacement of the redshift of DM-DE equality that affects the onset of the accelerated expansion (see Figure \ref{fig:Density}). An outstanding prediction of the present model is a negative EoS for DM (see ritgh panel of Figure \ref{fig:Pressure-omegaeffe}) that can lead to non-trivial phenomenological implications at the background level. In general, taking large and positive $s$ values the background cosmological behavior is quite similar to the one predicted for the $\Lambda$CDM model as it can be seen from Figures \ref{fig:Density} and \ref{fig:Pressure-omegaeffe}. 
Nevertheless, at perturbation level things can be very contrasting, leading, for instance, to the suppression of the power spectrum \cite{Velten:2014xca,Blas:2015tla,Barbosa:2017ojt}, which is quite desirable to alleviate the excess of power existing in the standard cold dark matter scenario. Dissipative pressure typically attenuates the growth of structures, so we expect to obtain similar results but the real impact must assessed for each viscous model. Hence, depending on how much the grow rate of perturbations is suppressed it can lead to a suitable phenomenological scenario. In the non-linear regime, dissipative DM can lead to a small power spectrum suppression due to large acoustic oscillations and diffusion damping \cite{Berezhiani:2003wj}, as well as strong suppression of the mass function, preventing thus the formation of halos at some halo mass threshold \cite{Foot:2016wvj}. 
A possible future work is to obtain, by a modified halo model of structure formation, the non-linear matter power spectrum for our new prescription to check its consistency with the observed power spectrum.

In addition, in order to investigate how this new parametrization behaves in the causal framework, a first physically acceptable step would be to use the truncated version of the IS full causal theory, which has been shown to be equivalent in expanding universe scenarios \cite{pavon1982covariant,zakari1993equations,Dissipativecosmology}.

As a final remark, we draw the attention that it is widely accepted that the EoS for DM is $\omega_{m}=0$ for the standard pressureless cold DM, but it can be different in non-standard cosmologies or modified $\Lambda$CDM scenarios,  like the one including dissipative viscous pressure Eqn.~(\ref{eqn:EoSDM}).  Within the latter scenario an effective and dynamical contribution to the effective EoS may appear, which modifies the pressureless feature of the DM fluid, as has been extensively investigated in the literature (see section II).  Moreover, it is worthwhile pointing out that in the large $s$-limit, our viscous model behaves similarly to the $\Lambda$CDM model at future times, that is $\omega_{m}^{\rm eff}\to0$, as can  be  appreciated in the right panel of Fig.~\ref{fig:Pressure-omegaeffe} for the case $s=2$. 

A natural question that arises from our main results is, can the discussed features provide a successful explanation for the current tensions due to the bulk viscosity effect? In other words, is there any observational signal of dissipative pressure in the cosmological data we may infer? This is the main issue we plan to tackle in a future work.\\

\textbf{Data Availability Statement}: No Data associated in the manuscript.


\section*{Acknowledgments}

G.G acknowledges financial support from Vicerrector\'ia de Investigaci\'on, Desarrollo e Innovaci\'on - Universidad de Santiago de Chile, Proyecto DICYT, C\'odigo 042031CM$\_$POSTDOC. G.P. acknowledges financial support by Dicyt-USACH Grant No. 042231PA. E.G. thanks to Vicerrectoría de Investigación y Desarrollo Tecnológico (VRIDT) at Universidad Católica del Norte (UCN) by the the scientific support of Núcleo de Investigación No.7 UCN-VRIDT 076/2020, Núcleo de Modelación y Simulación Científica (NMSC). A.R. and N.C. acknowledge Universidad de Santiago de Chile for financial support through the Proyecto POSTDOCDICYT, C\'odigo 043131 CM-POSTDOC.

\bibliography{biblio.bib}

\end{document}